\documentclass{aa}  

\usepackage[toc,page]{appendix}
\usepackage{graphicx}
\usepackage{txfonts}
\usepackage{natbib}
\usepackage{amsmath}
\usepackage{float}
\usepackage[colorlinks=true,urlcolor=blue,citecolor=blue,linkcolor=blue]{hyperref}
\usepackage{xcolor}

\usepackage{orcidlink}
\usepackage{soul}

\definecolor{indiagreen}{rgb}{0.07, 0.53, 0.03}

\newcommand{\kms}{\mbox{km~s$^{-1}$}}

\begin{document} 

   \title{Exploring self-consistent 2.5~D flare simulations with \texttt{MPI-AMRVAC}}

%   \subtitle{}

   \author{Malcolm Druett$^{1}$\orcidlink{0000-0001-9845-266X}
          \and
          Wenzhi Ruan$^{1}$
          \and
          Rony Keppens$^{1}$\orcidlink{0000-0003-3544-2733}}
          
\institute{$^{1}$Centre for mathematical Plasma Astrophysics, Department of Mathematics, KU Leuven, Celestijnenlaan 200B, B-3001 Leuven, Belgium
              \email{malcolm.druett@kuleuven.be}
             }

   \date{Submitted: \today}

% \abstract{}{}{}{}{} 

  \abstract
  %Context
  {Multi-dimensional solar flare simulations have not yet included detailed analysis of the lower atmospheric responses such as down-flowing chromospheric compressions and chromospheric evaporation processes.
  }
  %Aims
  {We present an analysis of multi-dimensional flare simulations, including analysis of chromospheric up-flows and down-flows that provide important groundwork for comparing 1D and multi-dimensional models.}
  %Methods
  { 
  We follow the evolution of an MHD standard solar flare model including electron beams, where localized anomalous resistivity initiates magnetic reconnection. We vary the background magnetic field strength, to produce simulations that cover a large span of observationally reported solar flare strengths.
  Chromospheric energy fluxes, and energy density maps are used to analyse the transport of energy from the corona to the lower atmosphere, and the resultant evolution of the flare. Quantities traced along 1D field-lines allow for detailed comparison with 1D evaporation models.}
  %Results
  {The flares produced by varying the background coronal field strength between 20~G and 65~G have GOES classifications between B1.5 and M2.3. All produce a lobster-claw reconnection out-flow and a fast shock in the tail of this flow with similar maximum Alfv\'en Mach number of $\sim$ 10. 
  The impact of the reconnection out-flow on the lower atmosphere and heat conduction are the key agents driving the chromospheric evaporation and "down-flowing chromospheric compressions". 
  The peak electron beam heating flux in the lower atmospheres varies between $1.4\times10^{9}$ and $4.7\times10^{10}$ erg~cm$^{-2}$~s$^{-1}$ across the simulations.
  The "down-flowing chromospheric compressions" have kinetic energy signatures that reach the photosphere, but at subsonic speeds, so would not generate sunquakes. 
  The weakest flare generates a relatively dense flare loop system, despite having a negative net mass flux through the top of the chromosphere, i.e. more mass is supplied downward than is evaporated upward. The stronger flares all produce positive mass fluxes. 
  Plasmoids form in the current sheets of the stronger flares due to tearing, and in all experiments the loop-tops contain turbulent eddies that ring via a magnetic tuning fork process. 
  }
%Conclusions
  {The flares presented have chromospheric evaporation driven by thermal conduction and the impact and rebound of the reconnection out-flow, in contrast to most 1D models where this process is driven by the beam electrons. 
  Several multi-dimensional phenomena are critical in determining plasma behaviour that are not generally considered in 1D flare simulations. These include loop-top turbulence, reconnection out-flow jets, heat diffusion, compressive heating from multi-dimensional expansion of the flux tubes due to changing pressures, and the interactions of upwards and downwards flows from the evaporation meeting the material squeezed downwards from the loop-tops.}

%\keywords{Line: profiles -- Methods: data analysis --
%   Techniques: polarimetric -- Techniques: spectroscopic -- techniques: miscellaneous -- Sun: atmosphere -- Sun: flares
%               }

   \maketitle
%
%________________________________________________________________
\section{Introduction} \label{sec:intro}

\begin{figure*}
    \centering
    \includegraphics[width=0.23\textwidth]{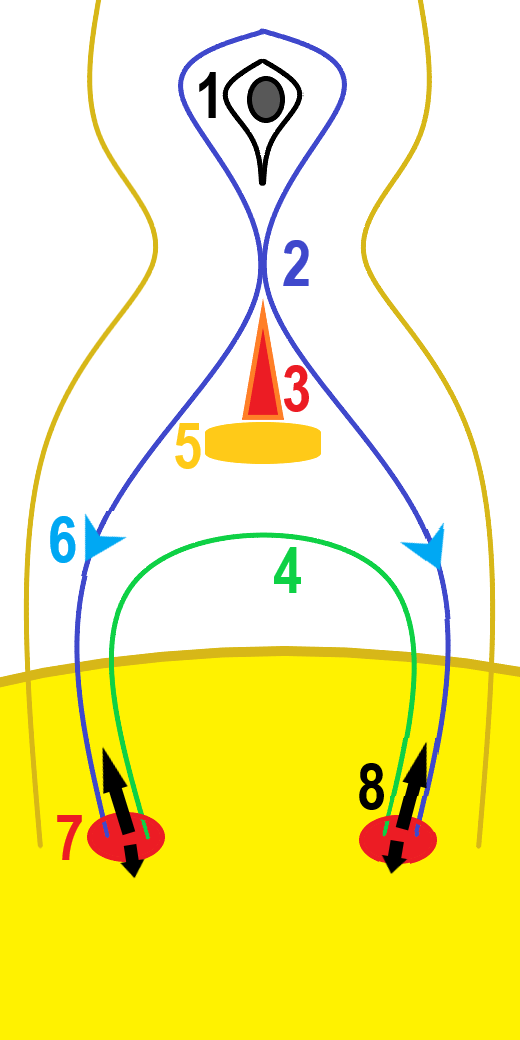}
    \includegraphics[width=0.73\textwidth]{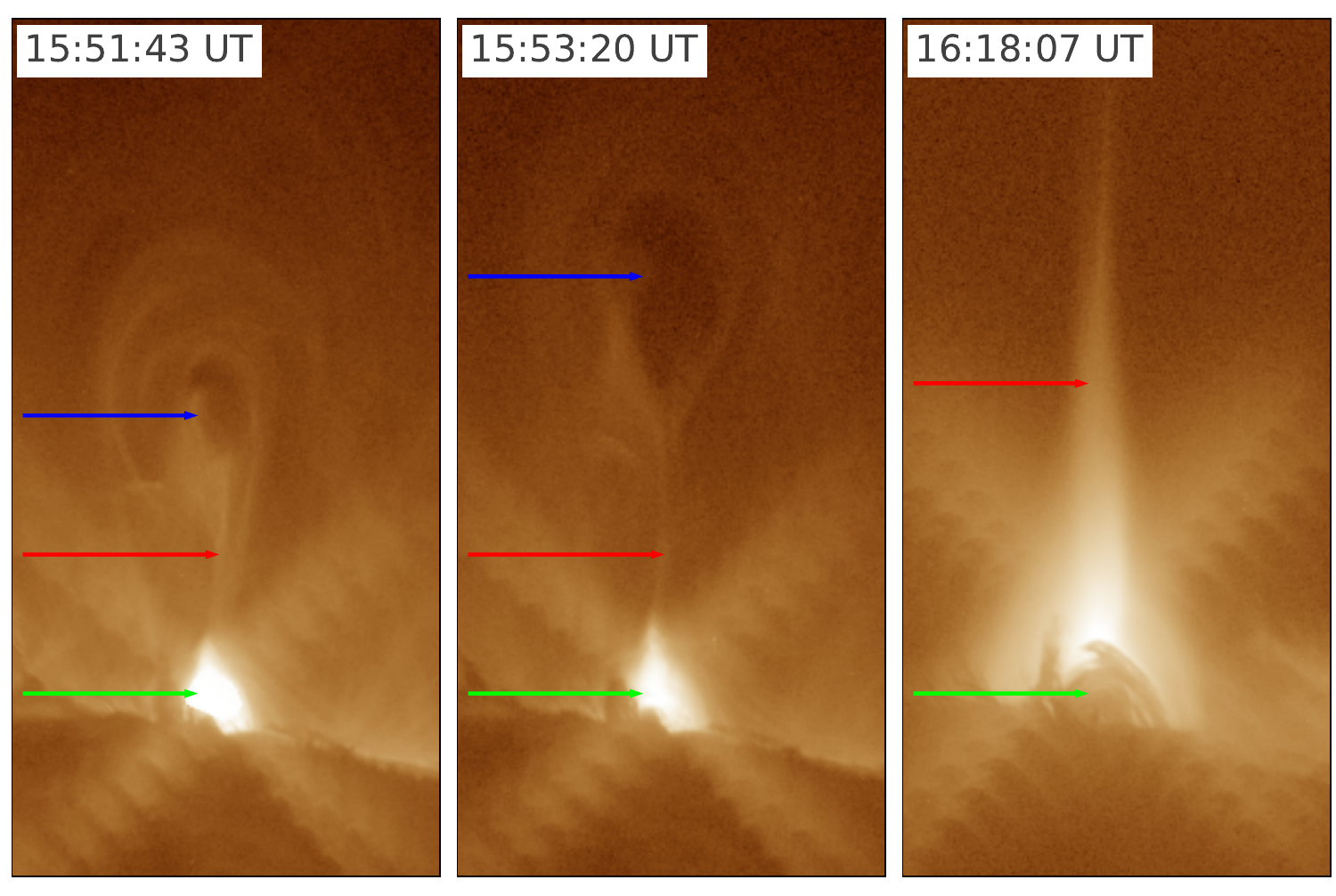}
    \caption{The standard solar flare model. \textit{Left}: The standard solar flare model in 2D, with features labelled as per the numbered points in section \ref{sec:intro}, namely (1) a flux rope running normal to the plane of the cross section, (2) reconnecting field-lines at an X-point, (3) reconnection out-flow jets, (4) reconnected hot loops that gather below the X-point, (5) a termination shock, (6) field-aligned transport of energy down to the lower atmosphere (7) flare ribbons, and (8) chromospheric evaporation and "down-flowing chromospheric compressions". \textit{Right}: Images of a solar flare over the solar limb taken with AIA aboard SDO \citep{2012LemenSDO} on Sept 10th 2017, showing the form of the standard solar flare model early in the flare (at 15:51:43 UT) in the 193 \AA\: channel with the loop arches at the base (green arrows) and a flux rope above it (blue arrows). Slightly later (at 15:53:20 UT) we see the flux rope erupting (moving upward). Later (at 16:18:07 UT) the flux rope has left the field of view and we see emission that is interpreted to originate from the long bright current sheet (red arrows), and dense coronal flare loops below (green arrows). These images use the short exposure AIA filters, but still also display significant fringe artifacts emanating outward from the bright loop arches.}
    \label{fig:AIAmodel}
\end{figure*}

In this paper we present multi-dimensional simulations of solar flares with a focus on the lower atmospheric dynamics that result from the coronal energy release.

% solar flare standard model intro
The first solar flare models based on magnetic reconnection \citet{1958Sweet, 1964Petschek} were developed in the mid 1900s \citep{1963Parker, 1964Carmichael, 1966Sturrock}, and led to the so-called standard CSHKP flare model \citep{1992SturrockCSHKP, 1996Shibata}. These models explained generalised observed behaviours of solar flares using the release of stored magnetic energy through magnetic reconnection events (see Fig.\ref{fig:AIAmodel}). 
This cartoon model has been expanded upon by many works, which are summarised well in \cite{2002PriestForbes}. Subsequent 3D models have stressed the role of current concentrations and quasi-separatrix layers \citep{2012Aulanier, 2013Aulanier, 2013Janvier, 2014Janvier}. 

% cross section of std flare model
The 2.5D simulations in this work present much simpler magnetic topologies than those in 3D models, but will account for all the relevant thermodynamic processes and the effects of electron beams.
Common features of such cross sections through solar flare models, are:
\begin{enumerate}
    \item A flux rope that runs perpendicular to the 2.5D cross section, and an underlying sheared flare loop arcade with a reconnection site in between \citep{1992SturrockCSHKP, 1996Shibata}. In eruptive flares this flux rope is ejected outward (upward).
    \item A magnetic x-point below the flux rope, where successive field lines are drawn inward, reconnected, and ejected outwards. 
    \item Fast reconnection out-flow jets from the x-point, in the orientation of the current sheet, one of which is directed towards the solar surface \citep{1964Petschek}. These shocks can take a lobster claw form during the initial ejection \citep{2011_Zenitani_reconnection_outflows}.
    \item Reconnected loops, ejected towards the surface, group together below the x-point and form hot coronal loop arcades \citep{1966Sturrock, 1996Shibata}.
    \item The out-flow jets become super-Alfv\'enic, and establish slow-mode shocks at their edges. A fast-mode termination shock forms at the interface between the core of the out-flows and the hot coronal loop arcade below \citep{1996Shibata, 2001Yokoyama, 2020RuanFlare, 2022Shen3DFlare, 2023Ruan3D}.
    \item Energy is liberated from the magnetic field at the reconnection x-point and potentially in a number of other larger volumes throughout the flare loop system, for example, in slow shocks in the reconnection jets \citep{1964Petschek, 2002PriestForbes}. The liberated energy is converted into many forms including heat, with models showing a significant fraction can be converted into the acceleration of high energy non-thermal particles \citep{2017Rowan, 2018Werner, 2023Hoshino}. Energy is transported from the neighbourhood of the x-point, and the hot coronal loop-tops, along magnetic field lines and towards chromospheric footpoints near the surface of the Sun. There are many proposed processes for this transport \citep[see review by ][]{2011Zharkova}. A key process is energetic particle acceleration (electrons and ions) along the field-lines \citep{1972Syrovatskii, 1978Emslie, 2011Holman, 2022Kong} with the acceleration mechanism falling into three broad categories: Direct Current electric field acceleration by an electric field above the Dreicer limit \citep{1959Dreicer, 1960Dreicer}, shock acceleration \citep{1985Ellison}, and turbulent/stochastic acceleration \citep{1996Miller, 2012Cargill, 2017Kontar}, including by the Kelvin Helmholtz Instability in turbulent loop tops \citep{2016FangFlareKHI, 2018RuanKHI}. Other energy transport mechanisms that will be present include thermal conduction predominantly parallel to the field-lines \citep{1962Spitzer, 1968Spitzer, 1969Spitzer}, and Alfv\'en waves \citep{2008FletcherAlfven}.
    \item Flare ribbons are the chromospheric locations where much of the released magnetic energy is deposited. This energy heats and excites the plasma and producing increased emission. Hard X-ray (HXR) sources are generally most intense in footpoints of the flare-loops. HXR sources are believed to result from bremsstrahlung of non-thermal (high-energy) electrons that lose their energy in collisions with the ambient thermal plasma. These energetic non-thermal distributions of particles are believed to accelerated near the x-point, and may also be accelerated in the termination shock and the turbulent reconnection that occurs in the tops of the loop arcades or in multi-stage acceleration processes \citep{2011Holman, 2011Zharkova, 2019KongTrappingTermination, 2020KongAcceleration}.
    \item The energy released in the flare loop foot-points causes hot up-flows (chromospheric evaporation) and cooler down-flowing "chromospheric compressions" towards the solar surface. The chromospheric evaporation fills the flare loops with hot dense plasma causing bright emission in the UV lines and soft X-ray (SXR) spectrum \citep{1969Bruzek, 1974HirayamaEvaporation, 1972Syrovatskii, 1985Fisher, 2016Polito, 2017Druett, polito_ribbons_2023}.
\end{enumerate}

% chromospheric compressions
In coronal physics a "condensation" refers to material that is dramatically cooling (with temperatures decreasing by an order of magnitude or so) and, typically, dropping from a higher ionisation state into a lower or partially unionised state. When this occurs the material suddenly becomes visible in chromospheric spectral lines, appearing to "condense".  "Chromospheric evaporation" refers to an up-flow of chromospheric material with simultaneous heating (temperature increasing by an order of magnitude or so), typically including a large increase in the ionisation degree of a particular state for an element. Although this "evaporation" is technically an ablation, the material appears to "evaporate" from chromospheric spectral lines. Therefore, we reserve the words "evaporation" and "condensation" for processes that have, at the very least, some semblance of a change in state. We note that the "down-flowing chromospheric compressions" in flares have often been referred to in the literature as "chromospheric condensations". However the conditions for the "condensation" analogy do not hold true for these phenomena. Therefore, we refer to this phenomenon as a "down-flowing chromospheric compression" for the rest of the paper (including quotation marks as a reminder to the reader).

% 1D modelling
The specialised methods of energy transport in this standard model (points 6-8 above) are a particularly challenging aspect for simulations. They have been investigated via 1D radiative transfer (RT) and hydrodynamic (HD) codes. These simulations generate hot up-flows from the chromosphere, primarily driven by energetic electrons and protons, thermal conduction, or combinations of these processes \citep{1985Fisher, 1987Canfield, 2005Allred, 2007Zharkova, 2015Allred, 2017Druett, 2018Druett, 2019Druett, 2023UnverferthPREFT}. 

% Other flare models
Recently multi-dimensional magnetohydrodynamic (MHD) models have investigated plasma flows in flares. \citet{2019Cheung_flare, 2023Rempel_flare} used sub-photospheric velocity driving to build up and release energy in the corona. \citet{2019KongTrappingTermination, 2020KongAcceleration, 2022Kong, 2022Shen3DFlare} have reproduced and interpreted supra-arcade down-flows in the corona immediately above the flare loops, with \citet{2020KongAcceleration, 2022Kong} inspecting the energetic electron acceleration and transport without yet transferring energy out of, and back into the MHD simulation. \citet{2020Kerr1D3D} used a set of 1D loop models to build up a 3D flaring volume. However, the main focus of multi-dimensional flare simulations has been on coronal dynamics. The investigation presented in this work includes a significant focus on the lower atmospheric dynamics and provide methods that, for the first time, enable clear comparisons of results from 1D and multi-dimensional modelling.

% Our lineage
There have also been attempts to reproduce the standard model in multi-dimensional MHD models. \citet{2001Yokoyama} initialised their 2D model as a weak bipolar magnetic field region within a vertically stratified approximation of solar conditions. They used anomalous resistivity to trigger reconnection above the polarity inversion line, which resulted in a loop arcade forming beneath. The combination of the reconnection out-flows impacting on the lower atmosphere and the thermal conduction front drive chromospheric evaporation from the footpoints of the loops. The resultant dense coronal loop arcade matches the general evolution scheme of a solar flare.

\citet{2020RuanFlare} self-consistently built on the model of \citet{2001Yokoyama} in 2.5D using the message passing interface-adaptive mesh refinement versatile advection code \citep[\texttt{MPI-AMRVAC}, ][]{2012KeppensAMRVAC, 2014PorthAMRVACSolar, 2018XiaAMRVACSolar, 2023KeppensAMRVAC3}, and expanded it significantly by using the Ohmic heating term in selected regions of anomalous resistivity as an energy reservoir to accelerate non-thermal electrons along field-lines. This energy is redistributed along these field-lines using analytical solutions for the 1D thick-target modelling \citep{1978Emslie} with remotely deposited energy subsequently re-interpolated onto the automated, block-adaptive grid. However, the agents causing chromospheric evaporation (hot up-flows of plasma with chromospheric densities into the corona) in the original paper are thermal conduction and the impact of the out-flows on the lower atmosphere, transported down the flare loop arcade. 

A companion paper to this study presents the first self-consistent multi-dimensional model of this kind reproducing chromospheric evaporation via energetic particle beams \citep{2023DruettEvap}. \citet{2023Ruan3D} presents a simulated flare in 3D using this modelling suite, but without including beam electrons, to study the formation of MHD turbulence in the flare loop-tops. 

In this paper we explore the 2.5D models including beam electrons described in \citet{2020RuanFlare}, in particular how variations of the coronal field strength affect the resultant coronal and lower atmospheric dynamics. This investigation provides the first solid basis for the comparison between 1D radiation hydrodynamic flare simulations and multi-dimensional flare modelling results. Thereby, we also lay groundwork to assess the need for the inclusion of critical field-aligned 1D physics to be built into truly multi-dimensional flare models.

\begin{figure*}[!ht]
    \centering
    \includegraphics[width=0.75\textwidth]{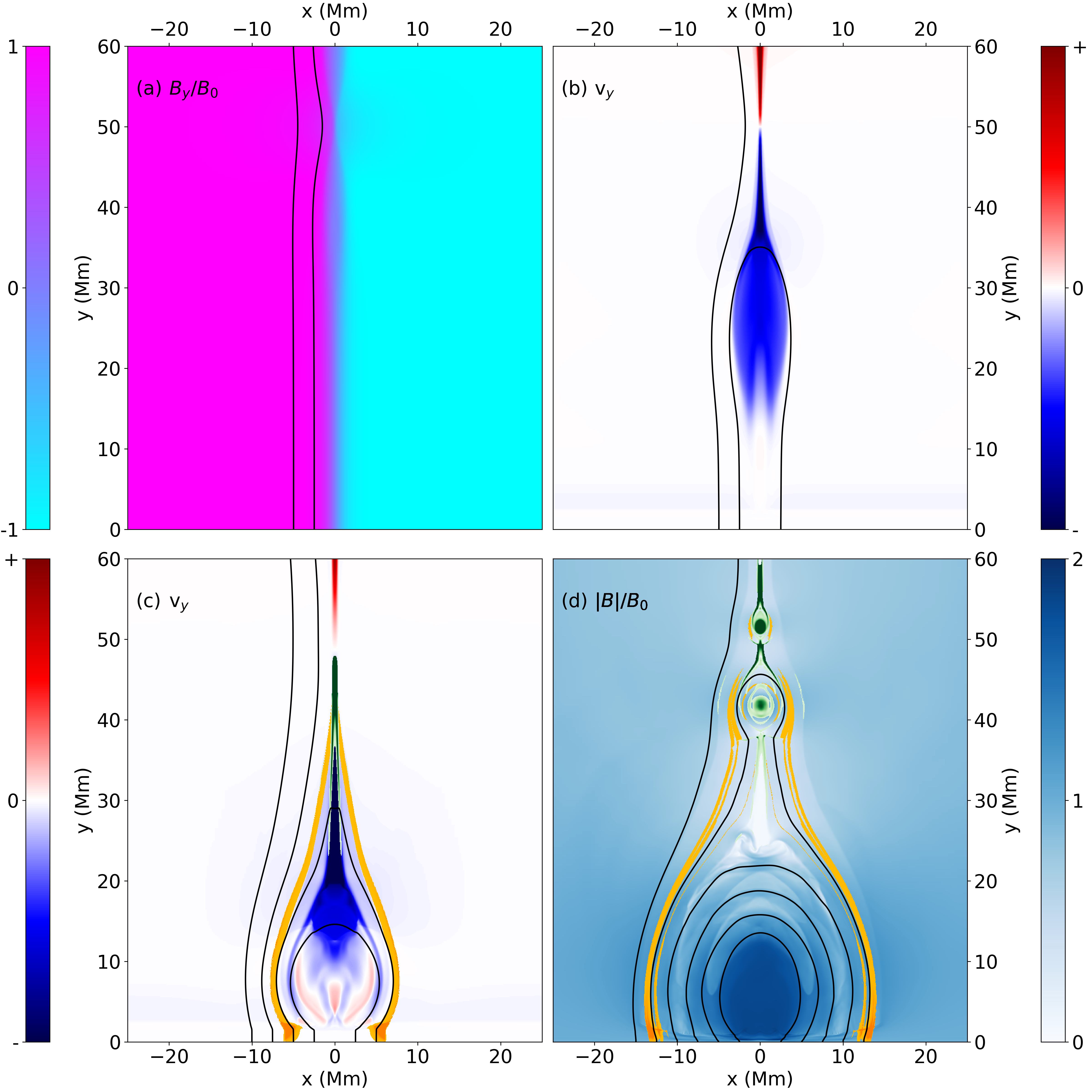}
    \caption{The simulated analogue of the standard solar flare model. Panel (a) shows $B_y / B_0$. All the simulations presented here start from a bipolar field region with an anomalous resistivity patch at a height of $y=50$~Mm, over the polarity inversion line, which causes the field to reconnect and release stored magnetic energy. Panel (b) shows the typical vertical velocity ($v_y$) structure at a time before the impact of the reconnection out-flow on the lower atmosphere. Positive velocity (red) represents upward motion and negative represents downwards motions (blue). The reconnection out-flow jets form from the X-point where reconnection occurs, one flowing downward and the other upward jet leaves to domain via the top boundary. Panel (c) shows the vertical velocity at the time when the reconnection jet impacts the lower atmosphere. Panel (d) shows the typical flare loop system that is formed via this process, through a plot of absolute magnetic field strength. Energetic electron transport is switched on after 31.2~s of the simulation. Thus, in all cases presented in this manuscript, the electrons are switched on before the impact of the reconnection jet on the lower atmosphere. On the lower panels, electron acceleration sites are shown in green, with energy deposition locations displayed in yellow. The energy deposition locations are saturated at very low values to help indicate their paths through the experiment. In fact, these beams deposit the majority of their energy in the lower atmosphere at the footpoints of the flare loops.} 
    \label{fig:StdModel}
\end{figure*}

\section{Model} \label{sec:model}
\subsection{MHD model description} \label{sec:equations}

The setup of the models in these experiment is comprehensively described in \cite{2020RuanFlare}. Here we only recall the equations used, along with an overview of how this simulation reproduces features of the standard solar flare model.
The MHD equations are, 
\begin{equation}
    \frac{\partial \rho}{\partial t} + \nabla \cdot (\rho \mathbf{v}) = 0\,,
\label{eq:cont}
\end{equation}
\begin{equation}
    \frac{\partial \rho \mathbf{v}}{\partial t} + \nabla \cdot (\rho \mathbf{v}  \mathbf{v} + p_{\mathrm{tot}}  \mathbf{I} - \mathbf{B}  \mathbf{B})= \rho \mathbf{g}\,,
\label{eq:motion}
\end{equation}
\begin{multline}
    \frac{\partial e}{\partial t} + \nabla \cdot (e \mathbf{v} + p_{\mathrm{tot}}  \mathbf{v} - \mathbf{B}  \mathbf{B}  \cdot \mathbf{v} ) 
    = \rho \mathbf{g} \cdot \mathbf{v} + \nabla \cdot (\kappa \cdot \nabla T) + \nabla \cdot (\mathbf{B} \times \eta \mathbf{J}) \\ 
    - Q_r - Q_e + H_b + H_e\,,
\label{eq:energy}
\end{multline}
\begin{equation}
    \frac{\partial \mathbf{B}}{\partial t} + \nabla \cdot (\mathbf{v}  \mathbf{B} -  \mathbf{B} \mathbf{v} ) = - \nabla \times (\eta \mathbf{J})\,.
\label{eq:induction}
\end{equation}

The equations are written in a dimensionless format. $\rho$, $\mathbf{v}$, $t$, $\mathbf{B}$, $e$ are the plasma density, velocity, time, magnetic field, and energy density.  ${\mathbf{g}}$ is the gravitational acceleration, which acts vertically downwards. This is calculated via the equation $\mathbf{g}= -274 R_{s}^{2} /(R_{s} + y)^2 \mathbf{\hat{y}}$ m~s$^{-2}$, where $R_s$ is the solar radius. $\mathbf{J}$ is the current density defined by $\mathbf{J} = \nabla \times \mathbf{B}$, and $\eta$ is the resistivity, with anomalous forms described in \citet{2020RuanFlare} and section \ref{sec:resistivity}.

Equation (\ref{eq:cont}) is the continuity equation, expressing the conservation of mass. Equation (\ref{eq:motion}) is the equation of motion also writable as, 
\begin{equation}
\frac{\partial \rho \mathbf{v}}{\partial t} + \nabla \cdot (\rho \mathbf{v}  \mathbf{v})= \rho \mathbf{g} - \nabla \cdot (p_{gas}) + \mathbf{J} \times \mathbf{B} \,.
\end{equation}
The Lorentz force $\mathbf{J} \times \mathbf{B}$ has been brought to the left hand side and decomposed into the magnetic tension and pressure making the total pressure $p_{\rm{tot}}=p_{\rm{gas}} + (B^2 /2)=p_{\rm{gas}}+p_{\rm{mag}}$. The form of these equations is discussed at length in section 4.3 of \citet{2010GoedbloedKeppensPoedts}.

Equation (\ref{eq:energy}) is the energy equation, where the total energy density is $e= (\rho v^2)/2  + p_{\mathrm{gas}}/(\gamma-1) + p_{\mathrm{mag}}$ with $\gamma = 5/3$. The first term on the right side of equation (\ref{eq:energy}) represents gravitational potential energy and additional sink or source terms on the right side express heat conduction (with a thermal conductivity tensor $\kappa$), resistive effects (The term $\nabla \cdot (\mathbf{B} \times \eta \mathbf{J})$ results from the inclusion of Ohmic dissipation, but is not Ohmic heating. Instead it shows that Ohm's law specifies the comoving electric field to be $\eta\mathbf{J}$, and that total energy remains conserved in resistive MHD, see section 4.4.2 of \citet{2010GoedbloedKeppensPoedts}), optically thin radiative losses ($Q_r$) and an artificial background heating that maintains the quiet sun coronal temperature (see equation 5 of \cite{2020RuanFlare}). 

In the second phase of the resistivity description, we take the Joule heating term $|\eta J^2|$ out of the local energy equation and use it as a reservoir of energy for the acceleration of energetic electrons (see section \ref{sec:beamdesc}). The term $Q_e$ represents this energy that is lost from the sites of energetic electron acceleration. The term $H_e$ represents the heating of the plasma by these energetic electrons, which is non-local and transferred along the magnetic field lines.

Equation (\ref{eq:induction}) shows the induction equation, which governs the advection of the magnetic field with the plasma. A source term on the right side describes the effects of resistive field diffusion, misaligned currents, and resistivity gradients (see section 4.4.2 and equation 4.132 of \citet{2010GoedbloedKeppensPoedts}). Coupled with the energy equation described above, this acts to convert magnetic energy into internal energy at sites of resistivity. The system of equations is closed by an ideal gas law as the equation of state.

\subsection{Anomalous resistivity description} \label{sec:resistivity}

The anomalous resistivity prescription in these simulations is a two-stage model. Both are calculated as described in \citet{2020RuanFlare} and were based on the earlier model of \citet{2001Yokoyama}. The first stage is used to trigger the reconnection at the x-point in the corona and takes the form
\begin{equation}
    \eta (x,y,t<t_\eta) =
        \begin{cases}
            \eta_0 \left[ 2 \left( \frac{r}{r_\eta} \right)^3 - 3 \left( \frac{r}{r_\eta} \right)^2 + 1 \right] , & r \leq r_\eta \\
            0 , & r > r_\eta .
        \end{cases}
        \label{eq:eta_1}
\end{equation}
$\eta_0=0.05$ is the maximum value of the anomalous resistivity in this stage. $r_\eta=2.4$~Mm is the radius over which the anomalous resistivity drops to zero, with distance $r$ away from the point ($0,50$)Mm.

The second phase is activated at times after $t_\eta>31.2$~s, which is $t_\eta=0.4$ units of experiment time,   
\begin{equation}
    \eta (x,y,t \geq t_\eta) =
        \begin{cases}
            0, & v_d > v_c \\
            \mathrm{min} \left\{ \alpha \left( \frac{v_d}{v_c} - 1 \right) \mathrm{exp} \left[ - \left( \frac{y-h_\eta}{h_s} \right) ^2 \right] , 0.1 \right\}, & v_d \geq v_c.
        \end{cases}
        \label{eq:eta_2}
\end{equation}

This generates anomalous resistivity in locations where the calculated electron drift velocity, $v_d (x, y, t) = \left( \frac{J}{e N_e} \right)$, is greater than a critical value $v_c = 1000 u_v$, where $u_v = 128$~\kms, $e$ is the electron charge, and $\alpha = 1 \times 10^{-4}$. The resistivity produced is given a maximal value of $0.1$. It is greatest at heights near the original resistivity patch, $h_\eta = 50$~Mm, and decreases with a scale height of $h_s=10$~Mm.

Ohmic heating results from the combination of the second stage anomalous resistivity description and the current in the current sheet through the domain about $x=0$. This Ohmic heating term, $\eta |J|^2$, is used to model unresolved microscopic instabilities, and in the second stage of our resistivity description is the energy reservoir $Q_e$ used to accelerate energetic electrons.
\begin{equation}
    Q_e (x,y,t \geq t_\eta) =
        \begin{cases}
            \eta |J|^2, & v_y < -u_v \,\,\,\mathrm{and}\,\,\, T > 5 MK \\
            0, & \mathrm{elsewhere}.
        \end{cases}
        \label{eq:Q_e}
\end{equation}

Any number of other resistivity schemes are available. These different schemes will impact the MHD evolution of the system. Investigation of large deviations from the scheme above is outside the scope of this work. It will be investigated in other works, including \citet{2023DruettEvap}.

\subsection{Beam model description} \label{sec:beamdesc}

The beam electron modelling used in this work is unchanged from the description of \citet{2020RuanFlare}, who describe the approach as a generalisation of the 1D treatment provided by \citet{1978Emslie}.

Firstly, we define the flaring region by tracing magnetic field-lines forward and backward from points to the left and the right of the reconnection x-point, at $(\pm 2.5, 50)$~Mm.
Within this flaring region field lines are traced from their photospheric footpoints. A check is made that neighbouring field-lines do not separate from each other by more than the separation of the grid-cell centres. If they do, additional field lines are added. The points where these field lines pass closest to the line $x=0$ are defined as the starting points for the subsequent 1D energy redistribution by beam electrons.

The input energy flux along each field-line is calculated as the total of the Ohmic heating terms of cells that intersect with each field-line, as described in Equation \ref{eq:Q_e}. The energy transferred to the field lines is subtracted from the MHD energy equation as $Q_e$. Where multiple field lines intersect a cell that produces energetic particles, the energy from that cell is shared equally between the field-lines.

Our numerical treatment of purely resistive MHD is fully conservative by construction as we use a finite volume approach, but partially open boundaries can modify exact conservation. Moreover, the additional presence of electron beams imply that at any instant, the beam injected energy is stored on (evolving) field lines, and given back to the plasma by interpolation from field lines to grid cells at a remote location. A consideration of how overall conservation is approximately achieved for the beam electrons at all times is given in Appendix \ref{sec:beam_heat_appendix}. 

\subsection{Domain and solution methods}

The equations in section \ref{sec:equations} are solved in a spatial domain spanning $-50$~Mm$<x<50$~Mm and $0$~Mm$<y<100$~Mm, using the open-source \texttt{MPI-AMRVAC} code \citep{2012KeppensAMRVAC, 2014PorthAMRVACSolar, 2018XiaAMRVACSolar, 2023KeppensAMRVAC3}. The hierarchical, block-adaptive grid used has a block size of 16 by 16 cells, with a minimum of 64 cells (4 by 4 blocks) spanning the domain in each dimension, at refinement level 1. The grid is refined by splitting a block into 4 sub-blocks for each increase of refinement level. The maximum refinement level for the experiment is 6. This means that at lowest refinement the grid cell separation is $100/64 = 1.5625$~Mm, and at maximum refinement the resolution is $100/64/2^5 = 48.8$~km. 

The refinement level is forced to be maximal below $y=3$~Mm, and for blocks within the box containing the dynamically tracked magnetic field-lines \citep[with the regions as described in Appendix B of][]{2020RuanFlare}. Additional automatic refinement and de-refinement is switched, on to ensure accurate shock capturing in locations away from the user-prescribed refinement areas. This is implemented with a weighting of 1:2:2 \citep[as described in][]{2012KeppensAMRVAC} between the conserved variables of mass density, vertical magnetic field, and internal energy density respectively.

We employ a three step time-stepping scheme. The flux scheme uses the HLL approximate Riemann solver \citep{1983HartenHLL} and as in \cite{2020RuanFlare}, a mixture of high-order slope limiters is used: a third-order limiter \citep{2009Cada} is employed in regions of low refinement, i.e. the background corona, and a second-order limiter \citep{1974VanLeer} in the regions of high grid refinement (greater than level 3), namely the lower atmosphere, reconnection region, and flare loop. The various limiters and all solvers available are discussed in \citet{2023KeppensAMRVAC3}.

\subsection{Differences to previous studies}

There a few differences between the experiments presented here and that produced in \citet{2020RuanFlare}, namely, 
\begin{itemize}
   \item The heat saturation model parallel to the field lines has been enacted using the \texttt{MPI-AMRVAC} thermal conduction module \cite{2018XiaAMRVACSolar}, with a monotonized central limiter as per \citet{1984Woodward}. 
   \item The side boundaries of the experiment ($x$-direction) are now treated as open, rather than the previously employed combination symmetric and asymmetric boundary conditions in the ghost zones. This avoids reflection of shocks that emanate from the flare, and these shocks simply exit the domain from the sides of the experiment. In \citet{2020RuanFlare} shocks reaching the side boundaries were reflected and returned and interact with the flaring region. As a result we now allow (negligible) mass loss in the chromosphere and the corona via the open boundaries over the duration of the experiment. The upper and lower boundaries are unchanged from \citet{2020RuanFlare}. 
   \item The magnetic field vectors are split. There is a constant background component with a distribution as in the model of \citet{2001Yokoyama}, and we solve for a time varying component that is zero at the start of the experiment. In \citet{2020RuanFlare} the background part of the field was given a magnitude of $B_0=35$G. In this investigation, 4 different values are used ($B_0= 20$G, $35$G, $50$G, $65$G) to explore the impact of different coronal field strength on the flare evolution. 
\end{itemize}

\subsection{How this experiment reproduces features of the standard solar flare model}

Figure \ref{fig:StdModel} shows how these experiments reproduce features of the standard solar flare model. The experiment is initialised with a low current-density vertical current sheet in the centre, where the vertical background magnetic field components transition from positive (left side of the experiment) to negative values (right side). This bipolar field region undergoes magnetic reconnection due to the anomalous resistivity region inserted at a height of 50~Mm. The current sheet thins and grows stronger in the reconnecting locations (Fig.~\ref{fig:StdModel}a). 

The reconnection, and associated expansion of the plasma due to heating, drives out-flows from the x-point. One of these out-flows is directed towards the surface of the Sun (see the blue patch in Fig.~\ref{fig:StdModel}b), and the other is directed upwards (the red jet in Fig.~\ref{fig:StdModel}b). Note that there is no overhead flux rope contained in this magnetic field configuration. We replicate only the portion of the standard solar flare model below the overlaying flux rope. In these experiments the magnetic field strength is chosen to reproduce realistic values in the corona, near the reconnection site, rather than at chromospheric and photospheric heights. Indeed, our maximal field values are on the order of $2 B_0$ in the lower corona. The chromospheric field strength values will be made more realistic in future works.
   
Electrons are accelerated from the reconnection site, in the out-flow regions, and around magnetic islands/plasmoids (see the green regions in Fig.~\ref{fig:StdModel}c and d). Such plasmoids are often caused by tearing events in thin current sheets in 2D simulations. The transport time for these electrons is considered to be shorter than the hydrodynamic time-steps of the simulation \citep{2009Siversky} and thus their energy transport is modelled as instantaneous. The energy deposition sites are shown in the lower panels using yellow colouration. This colouration is saturated at relatively low intensities to highlight the entire paths of the electrons, however, the energy deposition is actually focused in fairly concentrated kernels at the chromospheric foot-points of the flare loops. Particle trapping is possible in our beam model due to mirroring, and depends on the adopted beam pitch angle \citep{2020RuanFlare}. In such cases the energetic electrons remain on portions of the field-line in the next time-step of the simulation. In practice, the (yellow) beam visualizations of the energetic electrons act as a proxy to highlight the separatrices of the reconnected field-lines from those which are not currently reconnected. The inner regions of the electron energy deposition (i.e. $x$-locations closer to zero) are due to recently reconnected field-lines still accelerating electrons in the X-point out-flows or in plasmoids, but also shows loops that have retained some of their energy from earlier times due to trapping of energetic electrons.
Note that there are no specific mechanisms in these simulations to replicate particle acceleration in the termination shock or in turbulent flare loop-tops. However, this could be enacted in the future by judiciously generalizing the current heuristic recipes for the beams.

In these models the reconnection progresses rapidly from the start of the experiment, thus the out-flow jets impact somewhat directly on the chromosphere (see Fig.~\ref{fig:StdModel} c). In a pure-MHD (no beam electrons), but 3D model, \citet{2023Ruan3D} first initiated a gentle reconnection phase that led to the formation of a loop arcade before the impulsive phase began. The impulsive out-flow under these circumstances impacts first upon the loop-tops of this arcade before reaching down the field lines to the chromosphere. For ease of comparison to the experiments of \citet{2020RuanFlare} we replicate their set-up, resulting in a more direct impact of the out-flows on the lower atmosphere, and leave investigation of variations to a separate paper \citep{2023DruettEvap}. The impact of the reconnection out-flow on the lower atmosphere, and the heating of the lower atmosphere due to other processes such as thermal conduction, causes chromospheric evaporation. This is the heating of initially cool chromospheric material up to coronal values, and its associated expansion and up-flow into the coronal flare loops.

Material ejected downwards from the reconnection out-flows, as well as upflows from the chromospheric evaporation, increase the densities in the hot flare loops, and turbulence is also seen in the loop-tops below the termination shock (Fig.~\ref{fig:StdModel}d). Although the chromosphere is only treated in single-fluid, non-radiative MHD here, we will also inspect the downward propagating shocks in these models that are the equivalents of "down-flowing chromospheric compressions", and inspect the momentum they supply to the photosphere. We also calculate the SXR and HXR outputs, but present only a few relevant parameters for our analysis. The X-ray periodicity, light-curves, and other synthetic observables will be discussed in detail in a future work.

\subsection{Free parameters}

In the models presented here, there are several free parameters. The electron beams have energy profiles defined via a spectral index, lower cutoff energy, and initial mean pitch angle distribution. All of these are currently set to pre-determined values as in \citet{2020RuanFlare} ($\delta=4$, $E_c=20$~keV, and $\theta=18^{\circ}$ respectively) and will be updated to be based on relevant atmospheric parameters in a future work. 

The evolution of the model is also controlled by the description of the anomalous resistivity involving a switch between resistivity schemes described in \cite{2020RuanFlare, 2023Ruan3D}. Manipulation of resistivity parameters are presented in \cite{2023DruettEvap}, where we demonstrate how these can lead to electron beam-driven evaporation. 

The geometry of the flare is determined by the strength of the background magnetic field strength ($B_0$), the height of the resistivity patch, whether or not we insist on left-right symmetry, and thermodynamic values that initialise the atmosphere and the magnetic field structure. In this work we will focus on the variations of the background magnetic field strength, $B_0$.

\section{Results} \label{sec:results}

In section \ref{sec:out-flow} we discuss how the variations of the background magnetic field strength impacts the magnetic reconnection and out-flows. In section~\ref{sec:impact} we analyse the impacts on the lower solar atmosphere of the beam energetics (sec~\ref{sec:beam}) and the reconnection out-flows which has, to date, largely been overlooked in multi-dimensional flare simulations. We analyse down-flows and chromospheric compressions (sec~\ref{sec:impactdown}), up-flows and chromospheric evaporation (sec~\ref{sec:impactevap}). In section~\ref{sec:impactloops} we discuss the formation of the flare loop arcade and turbulence in the loop-tops. Section~\ref{sec:1D} analyzes 1D cuts along field-lines to investigate the evolution of the flare simulation in selected flux tubes. Much of the physics in flares is magnetic-field aligned, and there is a long history of detailed one-dimensional (1D) flare simulations. This section establishes a basis for the comparison of results in 1D with multi-dimensional research. In multi-dimensional experiments there is a diversity of flux tube configurations. Thus, in sections \ref{sec:x10}, \ref{sec:x12.5}, and \ref{sec:x15} we present the results for three such flux tubes with foot-points at $x=-10$~Mm, $x=-12.5$~Mm, $x=-15$~Mm respectively, to provide a more complete picture of the field-aligned physics.

\subsection{Reconnection and out-flow} \label{sec:out-flow}

\begin{figure*}[!ht]
    \centering
    \includegraphics[width=0.95\textwidth]{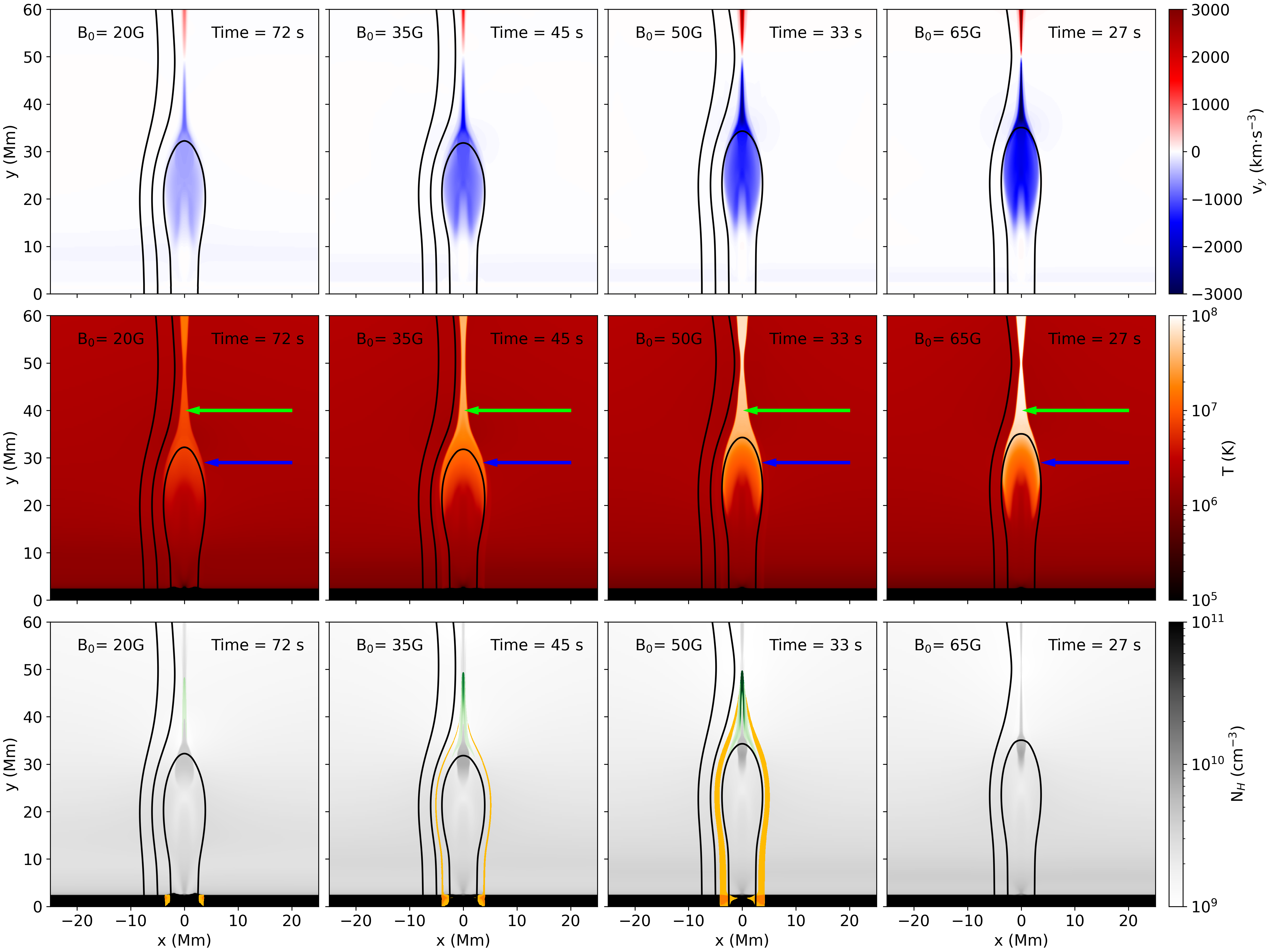}
    \caption{The reconnection out-flow jets of the experiments with different background magnetic field strengths. These are shown at similar morphological stages of the experiment evolution. The columns show results for different background magnetic field strengths from $B_0=20$~G on the left to $B_0=65$~G on the right. The top panels show the vertical velocities of the models, and the central row shows temperatures. In the temperature panels, green arrows indicate the concentration of high temperature in the tails of the reconnection out-flows, the blue arrows indicate the hotter areas at the rear of the lobster claw forms that lead the reconnection out-flows. The bottom row shows plasma number density. Each panel also shows magnetic field lines in black. These are traced from footpoints at $x=-2.5, -5.0,$ and $-7.5$~Mm in instances where these lines are being processed by the 1D field-line routines. In the number density panels (bottom row), beam electron acceleration sites are shown in green, and the locations where energetic electrons deposit their energy are shown in yellow.}
    \label{fig:outlfows}
\end{figure*}

\begin{figure*}[!ht]
    \centering
    \includegraphics[width=0.95\textwidth]{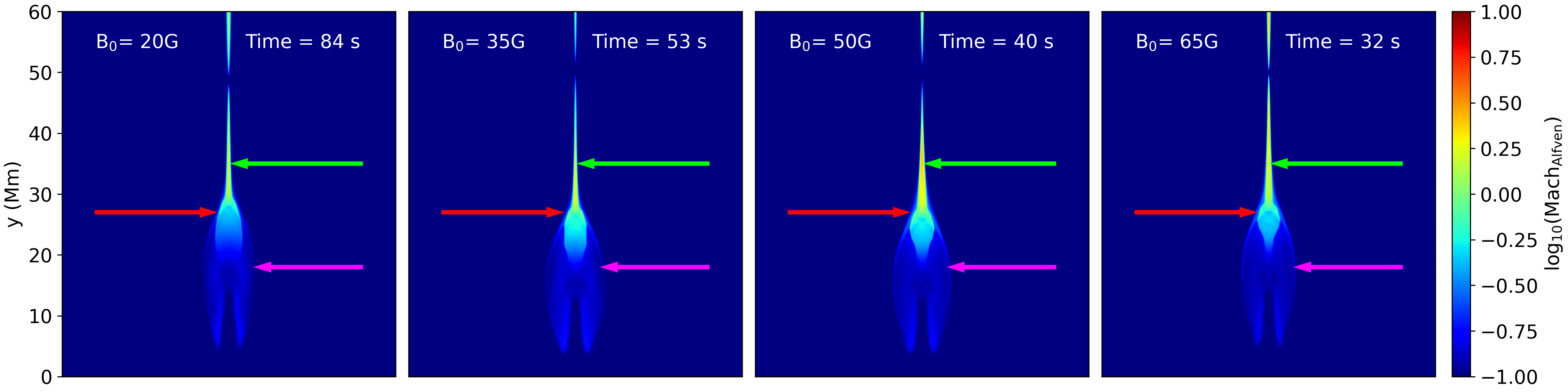}
    \includegraphics[width=0.95\textwidth]{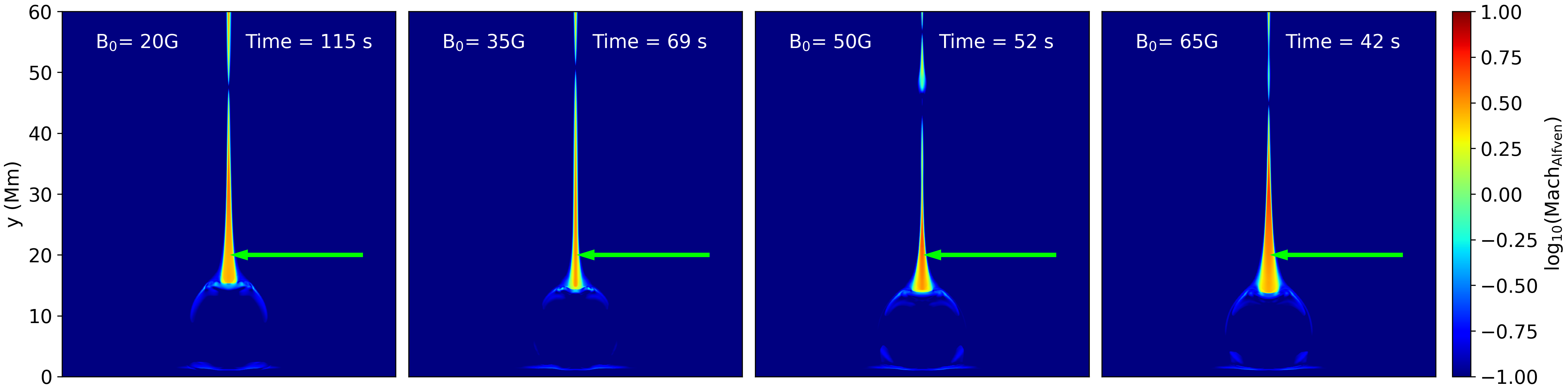}
    \includegraphics[width=0.95\textwidth]{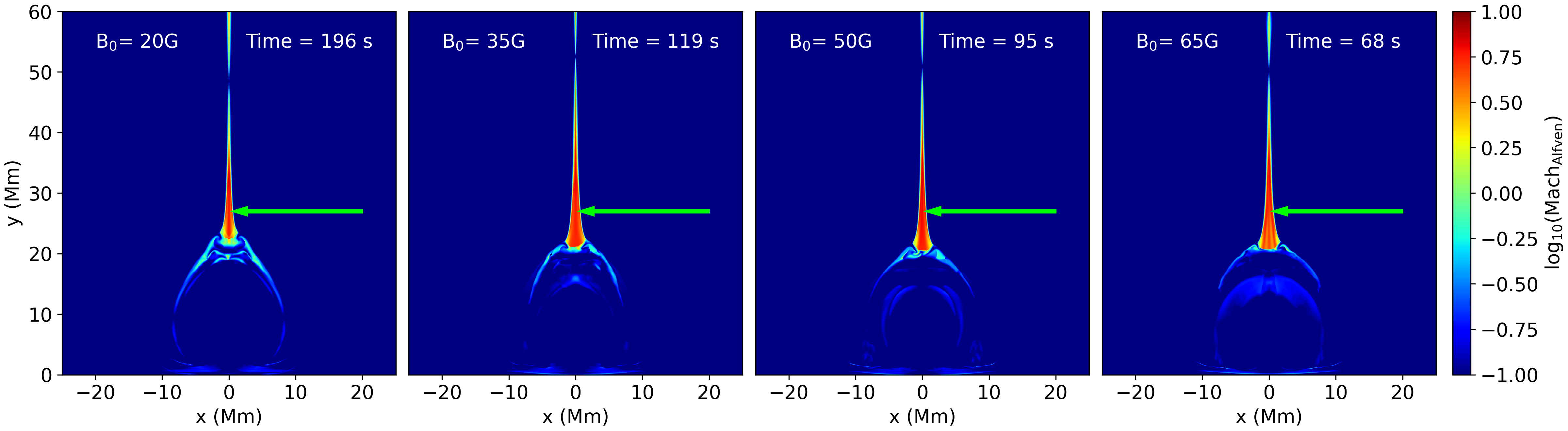}
    \includegraphics[width=0.95\textwidth]{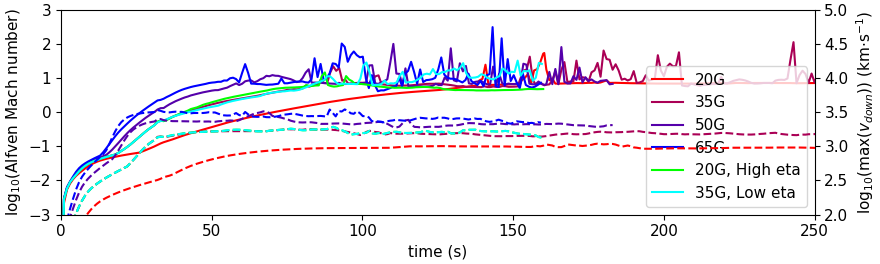}
    \caption{The Alfv\'en mach numbers of the out-flow reach similar ranges and values at similar evolution epochs, independent of field strength. These are shown just prior to the impact of the out-flow on the chromosphere (top row), during the compression of the flare loops by the generation of the termination shock (second row), and after the rebound of the impact once the flare loops have settled (third row). In each row a green arrow highlights the fast mode shock in the reconnection out-flow. In the top row a red arrow highlights the high-density core of the lobster claw out-flow formation, and a magenta arrow points to the sub-alfv\'{e}nic "claws" of this structure. In the lower panel the logarithm of the maximum downward out-flow speeds are shown with dashed lines (right axis), and the maximum of the Alfv\'en Mach numbers in these out-flows is shown with solid lines.}
    \label{fig:outlfowsalfven}
\end{figure*}

The initial atmospheres of each of the four experiments have identical thermodynamic variables. For the subsequent evolution, it is only difference in background magnetic field strengths that affects the release of magnetic energy.

The first hydrodynamic sign of the reconnection in each simulation is in the conversion of magnetic energy into internal energy along the reconnecting field-lines via Joule heating, which occurs before the electron acceleration process is switched on. By summing energy components over the entire domain in each simulation (not shown here for brevity) we see that the conversion of magnetic energy into internal energy continues relatively gently, while the conversion into kinetic energy accelerates with the development of the reconnection out-flow. Much of this kinetic energy escapes through the top boundary of the models or is re-converted into internal energy when the reconnection out-flow impacts the lower atmosphere. These down-flows also transiently and locally raise the magnetic energy when the out-flow compresses the magnetic loops down onto the lower atmosphere, before they rebound. 

The newly reconnected magnetic configuration, generates a Lorentz force. It is the combination of the heating and the suddenly altered pressure and Lorentz force values which drives the subsequent acceleration of the plasma away from the reconnection X-point. This out-flow (see Fig.~\ref{fig:outlfows}) forms a "lobster claw" shape for reasons discussed in \citet{2011_Zenitani_reconnection_outflows}, namely that in the fast-mode shock the density is highest in the central location and decreases away from the centre. This feature can be seen in the out-flow velocity plots (top row of Fig.~\ref{fig:outlfows}) with the velocity increasing with background field strength, due to the faster rate of energy release. The heating (second row, showing temperatures) is concentrated in the tails of these out-flows (see Fig.~\ref{fig:outlfows} central panels, green arrows), and to a lesser extent in the outer edges of the out-flows (blue arrows). The high density regions (bottom row) are concentrated in the central locations and some distance behind their leading edge. This core of high density material is more compact for simulations with stronger background magnetic field strengths (For context, see also the Alfv\'en Mach numbers of different sections of the "lobster claw" out-flow jets presented in Fig.\ref{fig:outlfowsalfven}).

Once the energetic electrons are activated in these models, the Joule heating energy term is removed from the energy equation and instead put into the acceleration of energetic particles in regions where the drift velocities of the electrons exceed a threshold value. In these locations the out-flows are still generated by the Lorentz force and other heating that results from the magnetic realignment and energy release, for example, shock heating and adiabatic compression (equation \ref{eq:energy}: $\nabla \cdot (p_{\rm gas} \mathbf{v})$ and thermal conduction (equation \ref{eq:energy}: $\nabla \cdot (\kappa \nabla T)$). 
 
In the bottom panels of Fig~\ref{fig:outlfows} the energetic electron acceleration regions (green) and energy deposition regions (yellow) are shown using a logarithmic colour-bars, so that the areas in which they are present is well highlighted. Again, these energy quantities are higher for the experiments with greater background magnetic field strengths, as the Joule heating term increases with the liberated magnetic energy. Note that we chose to visualize the 4 experiments in Fig.~\ref{fig:outlfows} at different experiment times, but at similar magnetic morphological times as seen in the selected field lines shown. Thus the panel of this figure showing the  $B_0=65$~G experiment does not have energetic beam electrons because they are switched on at $t=31.2$~s in all of the simulations. 

The dense core of the lobster claw shock accelerates towards the local Alfv\'en speed (see Fig.~\ref{fig:outlfowsalfven}, upper row, red arrows), with the claws travelling at significantly sub-Alfv\'enic speeds (see Fig.~\ref{fig:outlfowsalfven}, upper row, magenta arrows). Out-flows from the continuing reconnection increase in velocity to become a fast-mode shock. The fast shock forms in this tail of the out-flow (see Fig.~\ref{fig:outlfowsalfven}, green arrows). Independent of the background magnetic field strength, the out-flows reach a similar Alfv\'en Mach number of 9-10 in each simulation (see the solid lines in Fig.~\ref{fig:outlfowsalfven}, lower panel). 

To examine whether this maximal out-flow and Mach number is persistent, we also varied the free parameters that determine the anomalous resistivity values. Examples of the $B_0=35$~G experiment were run for 160~s of solar time with the anomalous resistivity a factor two greater and smaller values, and maximum threshold values as described in \citet{2020RuanFlare}, equation~(11) also increased or decreased by the same factor. Results of these experiments are included in Fig.~\ref{fig:outlfowsalfven}, lower panel and confirm that the limiting Alfv\'en mach number of the fast shock in the out-flows of the flare obtain similar maximum values independent of the resistivity or background magnetic field strength. Spiky behavior is due to the turbulent region overlapping with the diagnosed area, which was a fixed spatial box across all experiments, based on the typical region of the reconnection out-flow jet. Variations of the height of the resistivity patch, or asymmetries could also impact the maximum Alfv\'en mach number of the out-flow, which will be addressed in future work. 

The timing of the out-flow jet reaching this Mach number does not coincide with the timing of the maximum out-flow velocities reached in each experiment (compare the peaks in the solid and dashed lines in Figure.\ref{fig:outlfowsalfven}). In the stronger flare models the current sheet thins further and plasmoids form due to tearing instability, including a case of plasmoid coalescence in the experiment with $B_0=65$~G \citep[see e.g.,][]{2013Keppens, 2022SenPlasmoidTearing}.

\begin{figure*}
    \centering
    \includegraphics[width=0.95\textwidth]{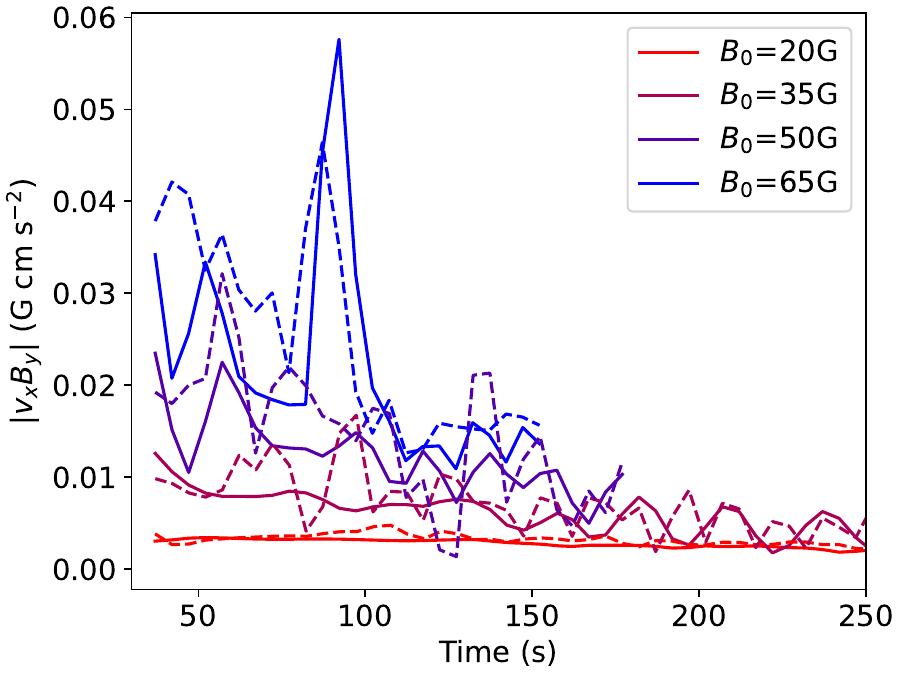}
    \caption{The reconnection inflow rate (solid lines) and footpoint sweeping (dashed lines) expressed as $|v_x B_y|$, in electric field units $\mathbf{E}  = -\mathbf{v}  \times \mathbf{B}$. The different colours show the variations in these quantities as functions of time for the experiments with different strengths of background magnetic field $B_0$.}
    \label{fig:reconrate}
\end{figure*}

The reconnection rate in the corona can be characterised using the ideal electric field, which is given by $-\mathbf{v}\times \mathbf{B}$ in the reconnection region and which, for the set-up used here, has a magnitude $|v_x B_y|$. \citet{2020RuanFlare} found that the sweeping of the footpoints in their simulations, located using the peak value in the footpoint HXR signal, related to this coronal reconnection flow via the relationship presented in \citet{2002Isobe}, $|v_x B_y|_{CORONA} = |v_{x(HXR)} B_y|_{FOOTPOINT}$.
Figure \ref{fig:reconrate} shows the values of $|v_x B_y|$ in the corona (solid lines) and $|v_{x(HXR)} B_y|$ shows values for the chromospheric footpoints (dashed lines). 
The units are converted into those of CGS ideal electric field to aid comparison with reconnection and acceleration studies which often use these units. We automated the calculations, in contrast to the previous hand-made calculations of \citet{2020RuanFlare} seen in their Fig.4. The reconnection inflow $|v_x B_y|_{CORONA}$ was calculated at $(-2,50)$~Mm. The values of $|v_x B_y|_{FOOTPOINT}$ were calculated from the $B_y$ values at the grid cell location of the maximum HXR emission in the chromospheric foot-point on the left side of the experiment, and $v_{x(HXR)}$ values were taken from the apparent horizontal motion. To account for the slow movement in terms of grid cell number in the footpoints, a ten-second moving average was used for the (signed) value of $v_{x(HXR)}$. Once the flare loop system is stabilised a clear periodicity of the measured reconnection rates appears in the simulation with $B_0=35$~G. This occurs in both the foot-points and the x-point, with the foot-point reconnection measure (about $10$~s) varying at half the period of the loop-top measure (about $20$~s). 

The relationship $|v_x B_y|_{CORONA} = |v_{x(HXR)} B_y|_{FOOTPOINT}$ holds much more consistently at later times in the experiment, once the flare loop system is formed, as was the case in \citet{2020RuanFlare}. It is in reasonable agreement across the range of background magnetic field strengths, $B_0$. This is consistent with \citet{2002Isobe}, which presented findings only after the initial phase of the flare, although we note numerous caveats for this comparison including the very different timescales involved.

The spikes for the $B_0=65$~G experiment at around $t=80-90$~s are produced by a passing plasmoid which increases the velocities and field strengths in the corona, as well as effecting the HXR footpoint locations in the chromosphere. The photospheric and chromospheric magnetic field strengths in our experiments are lower than reported solar flare field strengths by more than an order of magnitude. Typical solar magnetic field has a strong vertical gradient that is not present in our experiment. Thus, for more realistic flaring atmospheres one would expect much faster reconnection inflows in the corona than we find, if the footpoint sweeping speed was similar.

\subsection{Impact on the lower atmosphere} \label{sec:impact}

\subsubsection{Electron beam energetics} \label{sec:beam}

\begin{figure*}[!ht]
    \centering
    \includegraphics[width=0.95\textwidth]{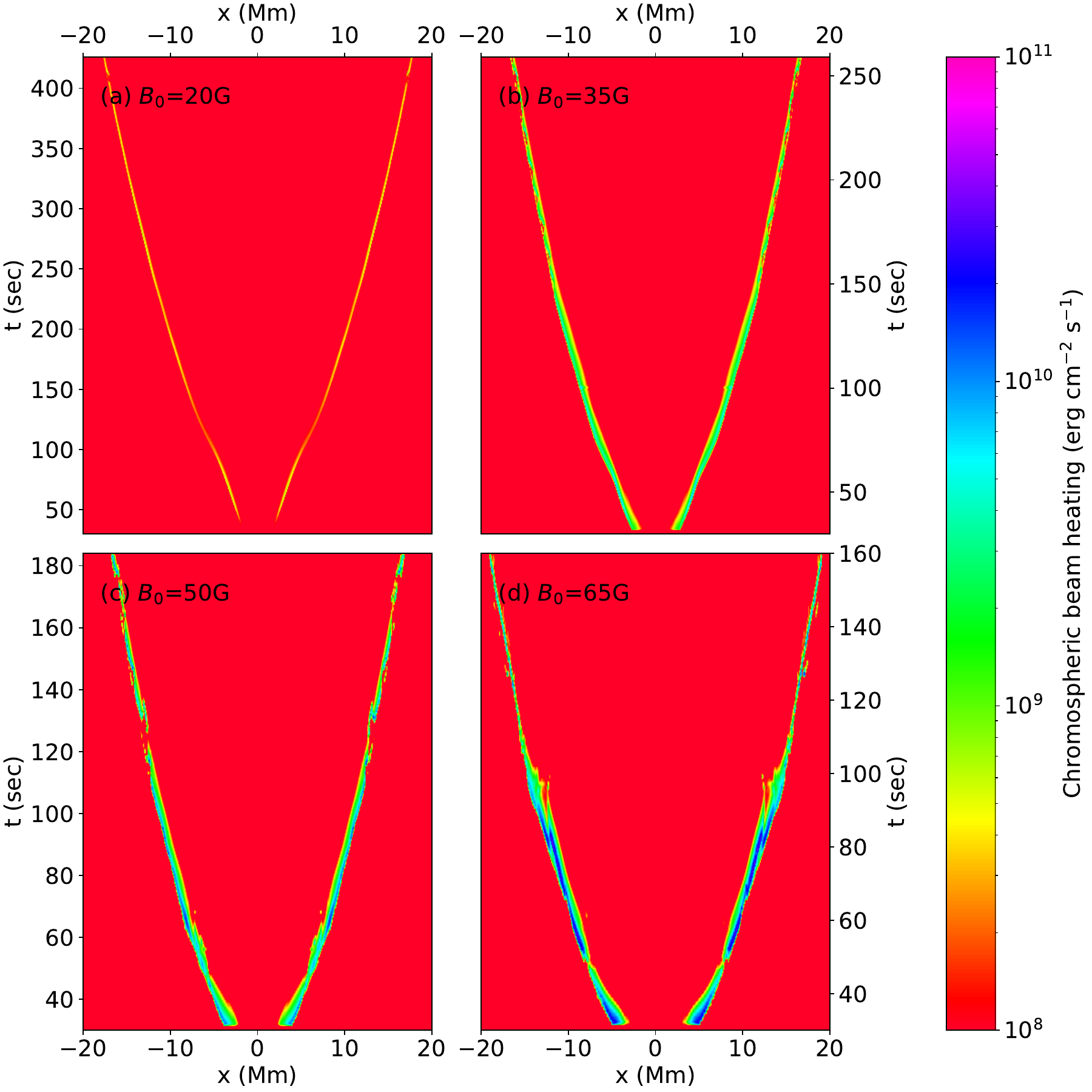}
    \caption{Electron beam heating in the chromospheric foot-points. Each panel shows results for an experiment with a different background magnetic field strength. These are shown as functions of time (y-axis), and footpoint location (x-axis). The heating at each footpoint is computed by integrating the source term for the electron beam heating rate. We integrate this quantity over a vertical distance in the spatial domain that spans from the lower boundary of the experiment up to (but not including) the grid cell where the temperature first exceeds 50,000K. The figure then shows the electron beam flux density that is applied down through the top of the "chromospheric" material and deposited at each footpoint. The colourmap saturates to red at the low end. This occurs at a beam strength of F8 ($F_0 = 10 \times 10^8$ erg cm$^{-2}$ s$^{-1}$), and thus the red colour indicates negligible or zero beam heating.}
    \label{fig:beamheat}
\end{figure*}

\begin{figure*}[!ht]
    \centering
    \includegraphics[width=0.8\textwidth]{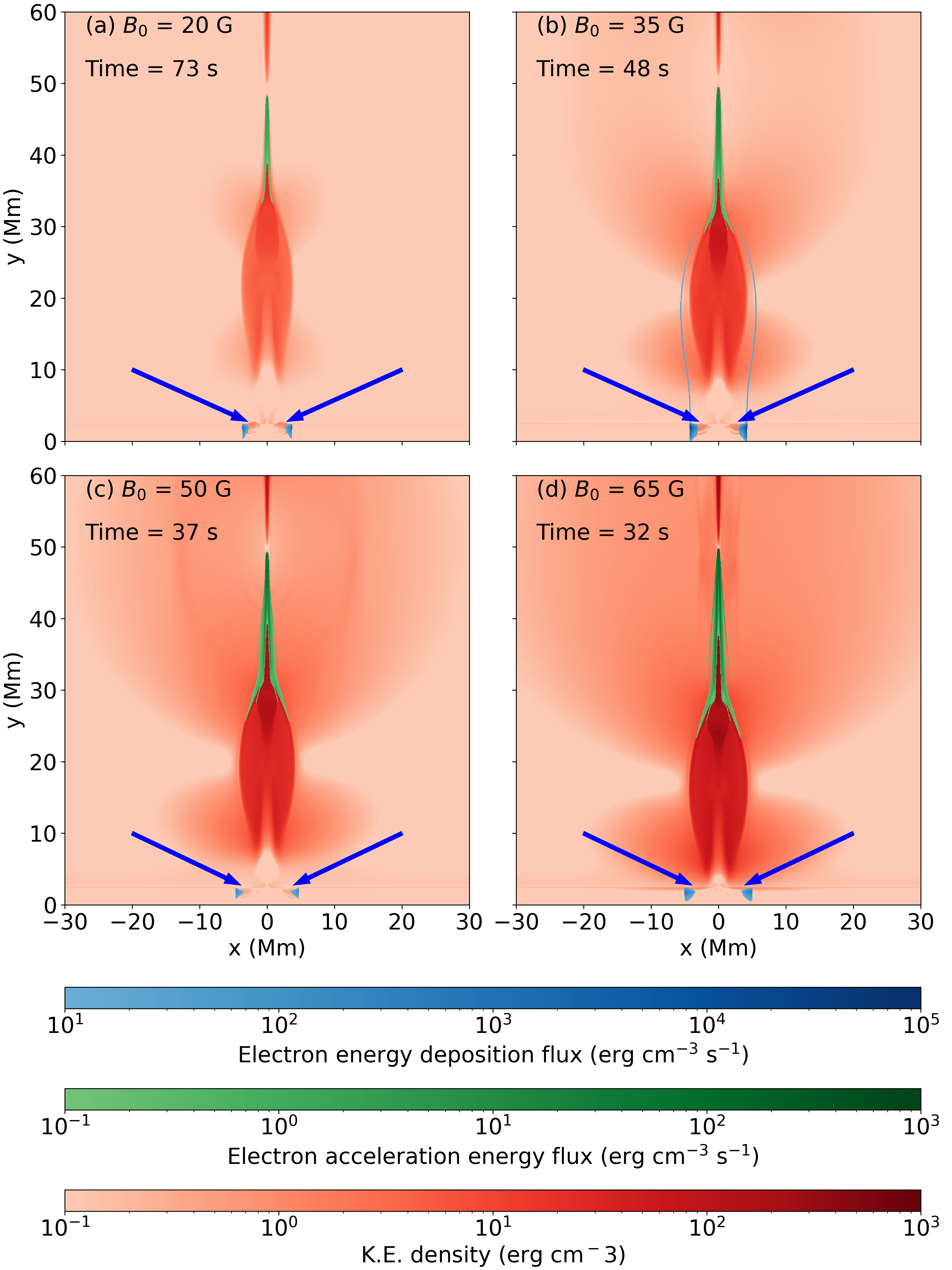}
    \caption{Signatures of the electron beam effects on the lower atmosphere. Kinetic energy maps of the flares are shown in red, with the lobster claw reconnection out-flows approaching the lower atmosphere for the experiments with $B_0$ values of (a) $20$~G at $t=73$~s, (b) $35$~G at $t=48$~s, (c) $50$~G at $t=50$~s, and (d) $65$~G at $t=32$~s. Overlays show the electron acceleration densities in green and the energy deposition regions in blue. Note that the lower panels have energy deposition rates masked and scaled to values 100 times greater than the upper panels, in order not to completely cover the footpoint kinetic energy signatures, which are seen as small red blobs (Kinetic energy) next to the blue footpoints blobs (electron energy deposition) and highlighted with blue arrows.}
    \label{fig:beamKEsignatures}
\end{figure*}

1D models of flares with energetic electron heated lower atmospheres generally do not use self-consistent energy schemes, instead injecting fluxes of high-energy electrons as functions of time at the tops of the models. The energies of these fluxes can be fixed to particular values or time profiles \citep{2005Allred, 2015Allred, 2018Druett, 2019Druett}, or can be driven by observational constraints \citep{2017Druett, polito_ribbons_2023}. Figure \ref{fig:beamheat} shows the chromospheric electron beam heating in our models, which can be compared with values used in 1D models like RADYN \citep{2005Allred, 2015Allred}, HYDRO2GEN \citep{2018Druett, 2019Druett}, and FLARIX \citep{2010VaradyFLARIX, 2017HeinzelFLARIX}. To compare a 1D beam model that uses a time-profile injected input heating with our multi-dimensional models, one should take a slice at a constant position (vertical slice) and read off the variations in footpoint heating flux. From the figure, one can see that our models have characteristic beam injection duration times of around 5-20 seconds, in line with some "elementary burst" models used in 1D simulations. 

It is clear that the model with $B_0=20$~G represents a very weak beam injection, with "F8" energy fluxes, i.e. an input energy flux $F_0$ on the order of $10^{8}$~erg~cm$^{-2}$~s$^{-1}$, peaking at values greater than $10^{9}$~erg~cm$^{-2}$~s$^{-1}$ at only a few locations within the domain. The $B_0=35$~G experiment is a reasonable analogue of a weak "F9" elementary burst model at most locations, although there is an absolute maximum flux value throughout the domain of $1.4\times10^{10}$~erg~cm$^{-2}$~s$^{-1}$. The stronger $B_0=50$~G and $B_0=65$~G models represent elementary burst models with duration of $5$ to $20$~s with moderate electron beam fluxes on the order of "F10", i.e. with $F_0 \approx 10^{10}$~erg~cm$^{-2}$~s$^{-1}$. On the basis of 1D modelling results in the literature one would expect the stronger flare models to produce some evaporation (hot up-flows) and cooler "down-flowing chromospheric compressions" signatures as a result of the beam electrons. Figure \ref{fig:beamKEsignatures} shows kinetic energy maps of the flare experiments at times after the electrons are switched on and before the impacts of the reconnection jets on the lower atmosphere. The kinetic energy signatures are shown in red with the electron energy deposition locations shown in blue. Before the impact of the reconnection out-flow jets there are indeed (minor) signatures in the red kinetic energy plots of up-flows and down-flows in these experiments (with locations indicated by blue arrows in Fig.\ref{fig:beamKEsignatures}). However, the reconnection out-flow jets which arrive and impact a bit later completely swamp these beam-driven evaporation signatures. The weaker flares have a much longer time window between the start of the beam heating and the impact of the reconnection jet, making it appear as if they have a greater influence on the lower atmosphere. When we instead look at similar times after the switching on of the beam, the stronger flares have stronger beams that exert a greater rate of influence, in line with what would be expected from their higher beam fluxes. In a separate paper we adapt the models presented here to investigate chromospheric evaporation driven primarily by electron beams \citep{2023DruettEvap}.

\begin{figure*}[!ht]
    \centering
    \includegraphics[width=0.95\textwidth]{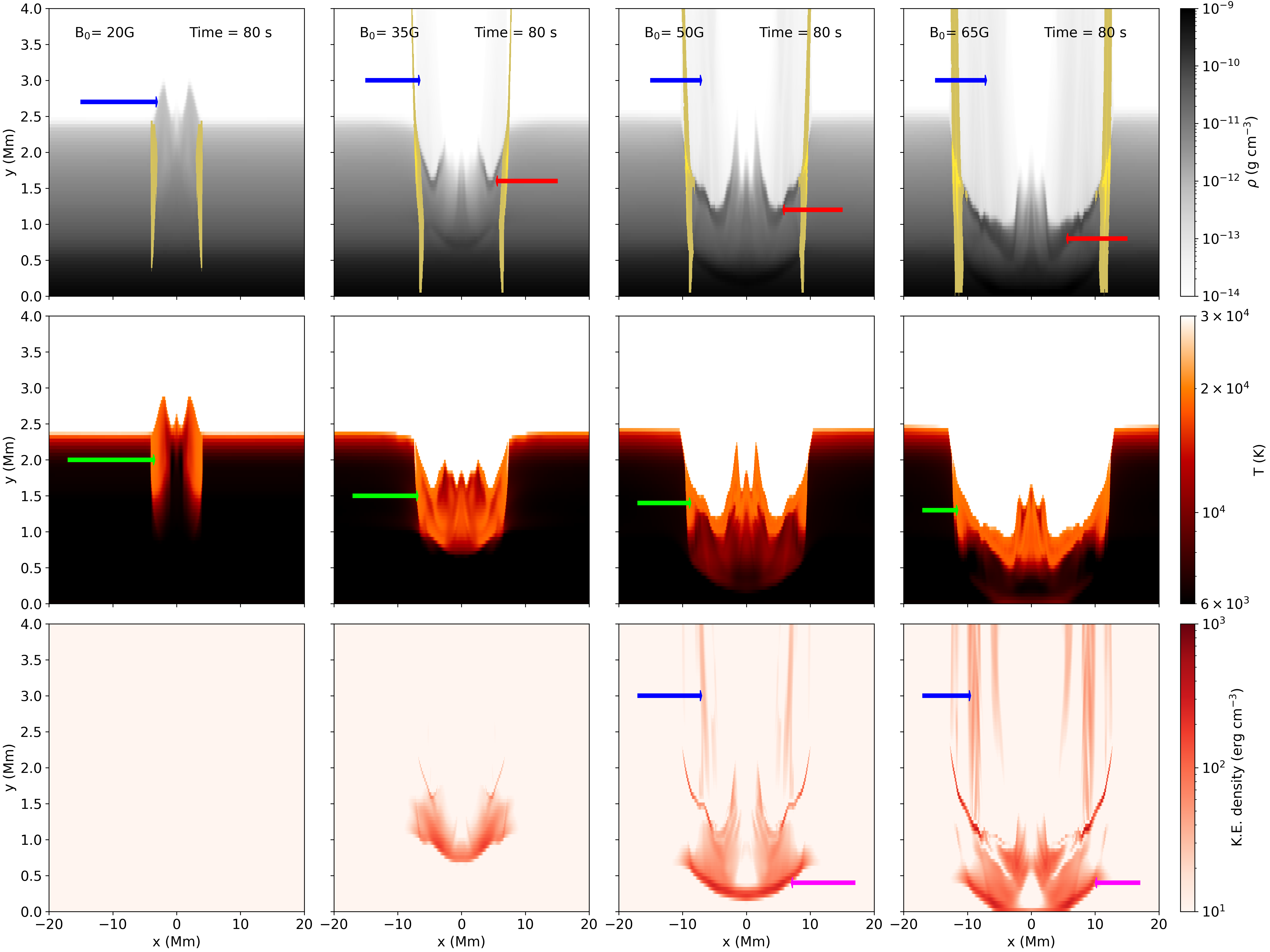}
    \caption{The heating, down-flows, and evaporation of chromospheric material. These are shown for simulations with increasing background magnetic field strengths in the columns from left to right, each at the same time during the simulation ($t=80$~s). The top row shows the plasma density in grey-scale, with electron energy deposition sites overlaid in yellow. Blue arrows indicate the locations of up-flows from the chromosphere, evaporation in the case of all but the $B_0=20$~G experiment. Red arrows indicate the high density impact fronts of down-flowing material at the top of the chromosphere. The central panels show the temperature of the atmosphere, saturating to black at 6000K and to white at 30,000K. In this row green arrows indicate hot chromospheric material. The bottom rows show the kinetic energy densities with chromospheric evaporation signatures highlighted using blue arrows and significant energy transfer to photospheric levels indicated by magenta arrows.}
    \label{fig:chromo80}
\end{figure*}
\begin{figure*}[!ht]
    \centering
    \includegraphics[width=0.95\textwidth]{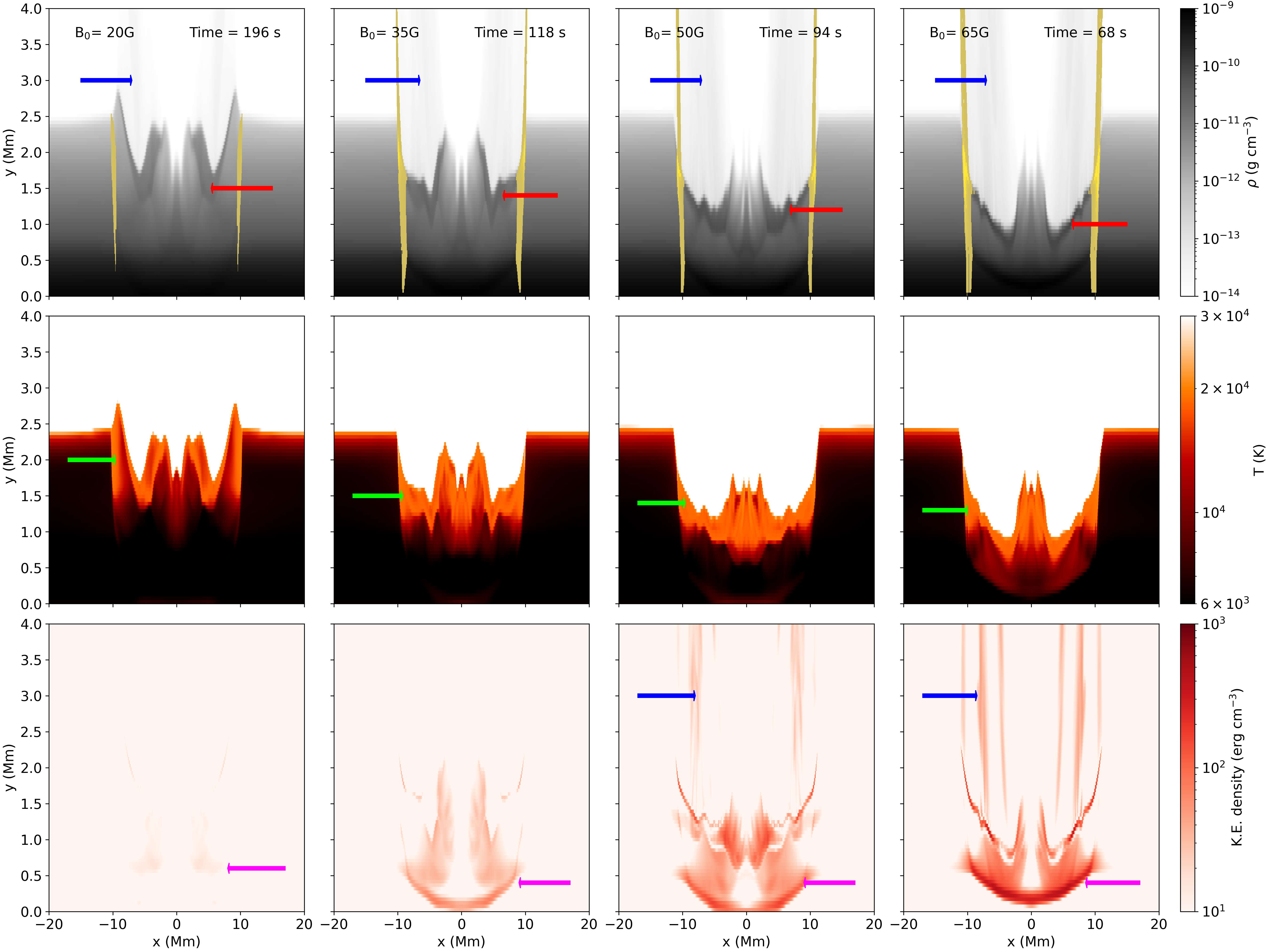}
    \caption{The heating, down-flows, and evaporation of chromospheric material in the models. The formatting is similar to that of Fig.~\ref{fig:chromo80}, including those features that are highlighted by arrows. This figure shows the simulations with different background magnetic field strengths, each at similar stages in the evolution of the flare, after the impact and rebound of the reconnection out-flow jet.}
    \label{fig:chromo_late}
\end{figure*}
\begin{figure*}[!ht]
    \centering
    \includegraphics[width=0.95\textwidth]{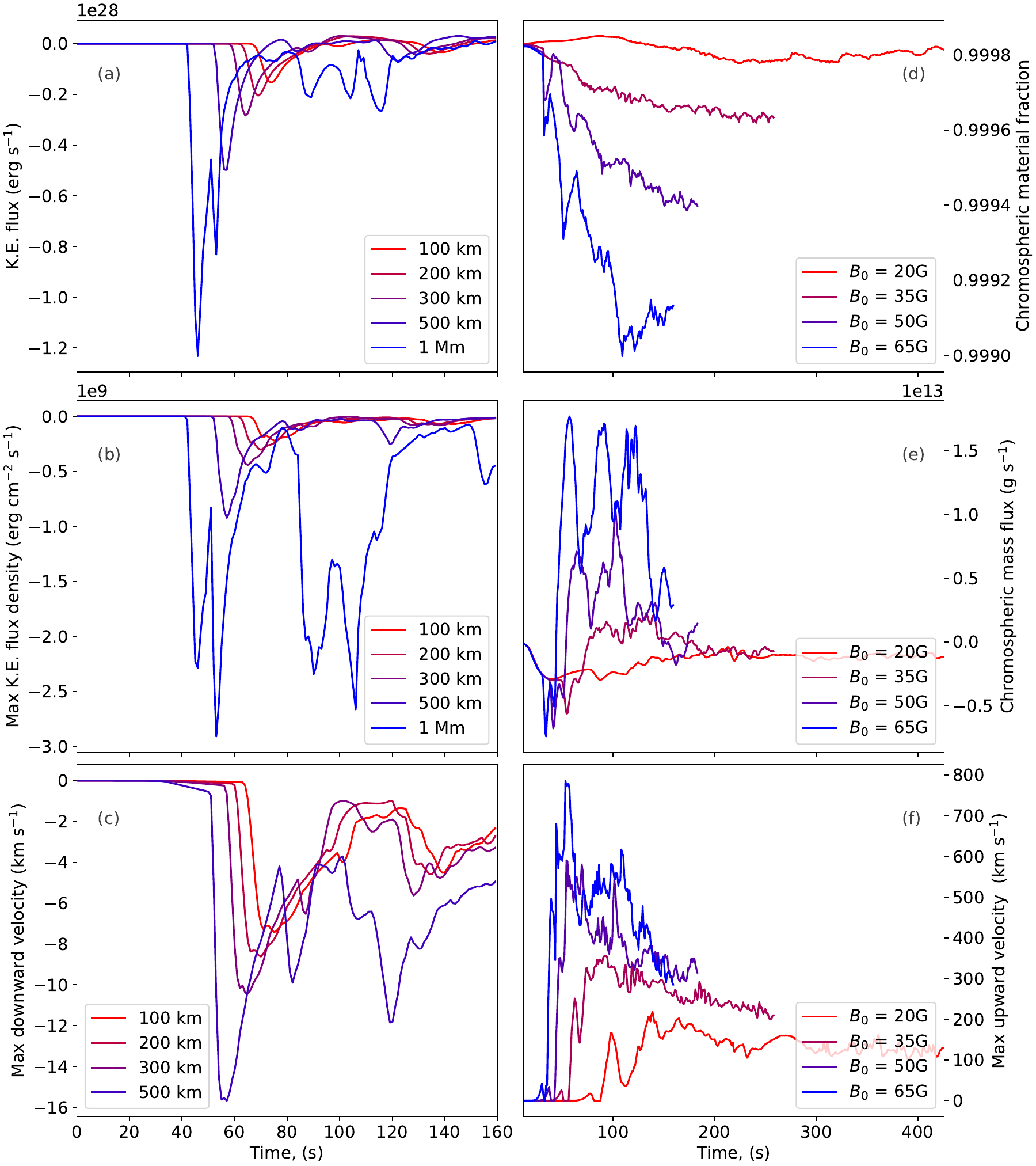}
    \caption{Impact of the flare on the dynamics of the lower atmosphere. The left column shows results for the $B_0=65$~G simulation, while the right panels compare all four experiments. Shown are: (a) the total kinetic energy flux (assuming a 3rd dimension depth of 100~Mm) (b) the maximum kinetic energy density directed downward, and (c) the maximum downward velocity, all shown at various heights near the photosphere. In the right column of panels we show: (d) the fraction of the lower atmospheric material that is "chromospheric" as functions of time for the different experiments, (e) the mass fluxes (assuming a 3rd dimension depth of 100~Mm) and (f) the maximum upward velocities of material, each taken at 5~Mm height.}
    \label{fig:downshocks}
\end{figure*}

\subsubsection{Chromospheric down-flows} \label{sec:impactdown}

In our MHD models the lower atmosphere is highly simplified. It is treated as a fully-ionised hydrogen plasma with a simple radiative loss function. The photospheric field strengths are of order 50~G, in rather stark contrast to the typical observationally derived values of a flare's lower atmospheric field strength, which are several kiloGauss. This seems to be a common situation for flare simulations derived to model coronal conditions, for example \citet{2020RuanFlare,2023Ruan3D, 2022Shen3DFlare}. Simulations developed originally from photospheric models that have been extended to accurately reproduce coronal conditions do not have this proviso, for example \citet{2019Cheung_flare, 2023Rempel_flare}. We shall refer to the low temperature, high density lower atmosphere region as the chromosphere despite its simplicity in our models, and to the region at the very base of our model as the photosphere, although we do not accurately reproduce this region of the Sun.

The instant the energetic electrons are switched on, they reach the lower atmosphere, as per the modelling assumptions. In all but the weakest field strength modelled, $B_0=20$~G, there are electrons that reach our photosphere, i.e. the base of the model, see Figs.~\ref{fig:chromo80} and \ref{fig:chromo_late}, top panels. The beam model used here, when implemented in 1D models, generally results in electrons being stopped at greater heights in the chromosphere \citep{1978Emslie, 2005Allred, 2015Allred}. The electrons do not impart directed momentum on the plasma in these simulations, acting only through a source term in the energy equation~(\ref{eq:energy}). 

Before the reconnection jets impact the chromosphere, the electron beams already heat the chromospheric footpoints from initial $\sim 6000$~K temperatures up to $\sim 20,000$~K over a range of heights that extends down to around 1.5Mm, in the $B_0=20$~G experiment (Fig.~\ref{fig:chromo80}, left panels). For the $B_0=65$~G experiment there is heating of plasma by the beam electrons to $T \sim 50,000$~K just above 2~Mm, and to $T \sim 20,000$~K at heights as low as 1~Mm. Moreover, for the $B_0=65$~G experiment this occurs before the impact of the reconnection out-flow arrives, despite the only 2 to 3 seconds delay between the switching on of the electron beams and the arrival of the reconnection out-flow jets (see $B_0=65$~G experiment at $t=34$~s in video form of the Fig.\ref{fig:chromo_late}). However, the beam heating does not cause significant up-flows in any of the experiments presented here. In the stronger flares there is not significant time for up-flows to form before the impact of the reconnection out-flow jet. In weaker flares the heating and expansion of the plasma has time to cause a gentle chromospheric up-flow (with chromospheric densities), reaching up to around 3~Mm (Fig.~\ref{fig:chromo80}, top-left panel, indicated by a blue arrow), but this does not qualify as chromospheric evaporation as it does not rise above this height.

The different stages of the down-flows in the lower atmosphere can be seen for each model at $t=80$~s in Fig.~\ref{fig:chromo80}. In the left panels (weak $B_0$) we see a lower atmosphere after the beam electrons have started heating it, but before the reconnection out-flows impact it. When the out-flows from the reconnection do impact the lower atmosphere they transfer downward momentum and kinetic energy, as well as increasing the pressure. There is conduction of thermal energy along field-lines due to the temperature gradient. These processes heat the lower atmosphere and push it downwards. Figures~\ref{fig:chromo80} and \ref{fig:chromo_late} (top panels) show dense impact fronts at the top of the hot flare chromospheres highlighted with red arrows. Above the flare chromosphere, in simulations with stronger background magnetic field, there is a stronger conversion of chromospheric material to hot plasma that up-flows into the coronal loops. This will be discussed in the next section, and these up-flows can be seen as grey patches of increased number density in the coronas of the top panels. They are also visible as coronal kinetic energy signatures in the lower panels, with both signatures indicated with blue arrows in the figures.

Meanwhile, below, the down-flow starts to slow and cool as it travels to the photosphere (Fig.~\ref{fig:chromo80}). Some downward travelling material is heated up to around $\sim 20,000$~K (green arrows) and below this there is very significant kinetic energy that travels down to the photosphere (magenta arrows). The left panels (a-b-c) of Fig.~\ref{fig:downshocks} shows the downward fluxes of kinetic energy, the maximum kinetic energy density and the maximum downward velocities at different heights through the atmosphere of the simulations with $B_0=65$~G. These heights were chosen at least five grid-points away from the experiment lower boundary, to avoid significant influence from boundary effects. The peak of the downward kinetic energy flux at a height of 300 km above the photosphere across the simulations with different $B_0$ values ranges from $2\times 10^{25}$ erg s$^{-1}$ ($B_0=20$~G) to $3\times 10^{27}$ erg s$^{-1}$ ($B_0=65$~G) with peak flux densities from $5\times 10^6$ erg cm$^{-2}$ s$^{-1}$ to $4\times 10^8$ erg cm$^{-2}$ s$^{-1}$. These fluxes were essentially unchanged across in simulations where we kept $B_0$ constant and varied the initial mean pitch angles (not presented here), due to the relatively low energy fluxes achieved via the energetic particles.

The "down-flowing chromospheric compressions" travel initially as acoustic shocks (at speeds greater than the sound speed just below them, which has typical values of 8-10 km s$^{-1}$). The "down-flowing chromospheric compression" in the $B_0=20$G flare ceases to be a shock in the mid chromosphere, when its downward velocity drops below 8 km s$^{-1}$. This transition occurs deeper into the model for increasing $B_0$, but even in the strongest flare, the compression is travelling below the sound speed by the time it reaches a height of 200-300~km above the photosphere. Therefore, the "down-flowing chromospheric compressions" in these simulations would not be expected to cause a sunquake \citep{2001Kosovichev_sunquake, 2018Macrae_Sunquake, 2020ZharkovaII} when they move below the photosphere.

\subsubsection{Evaporation} \label{sec:impactevap}

Fig.~\ref{fig:chromo_late} illustrates that the area of the chromosphere that gets heated, compressed, or evaporated due to the flare is greater with increasing field strength, both in depth (due to the higher energy fluxes) and in lateral area (due to the faster speed of the leading edge of the flare ribbon that results from the faster reconnection rates). The fraction of the mass at heights between 300 km and 5 Mm that is at temperatures less than 20,000K is quantified over time in Fig.~\ref{fig:downshocks}d, alongside the chromospheric mass flux through a horizontal line at 5 Mm through the experiment (Fig.~\ref{fig:downshocks}e). These values vary co-temporally, and don't show any in-phase behaviours with times of large photospheric mass fluxes. 

Due to the mass density gradient with height that occurs between the photosphere and the chromosphere, the general plasma motions in our simulations have larger mass fluxes through the bottom boundary than through the tops of the chromosphere. These are especially larger when the front from the chromospheric compression reaches the lower boundary. These fluxes were checked and the chromospheric mass fraction in panel d of Fig.~\ref{fig:downshocks} showed in-phase variations with the chromospheric mass flux but no in-phase variations with these photospheric fluxes. Thus, we can be confident that the evolution of chromospheric material in panel d is due to chromospheric rather than photospheric changes.

The mass flux via evaporation (panel e), and the decrease in chromospheric material over time (panel d), both increase with background magnetic field strength between $B_0=35$~G and $B_0=65$~G. In the case of $B_0=20$~G we see that the net mass flux is at first consistently downward from the corona to the chromosphere, and the change in the fraction of material at chromospheric heights that has temperatures greater than 20,000 K varies relatively negligibly. Thus, it is shown to be possible for high density flare loop systems to form that are fed primarily from the coronal reconnection jets and not from the chromospheric footpoints, since dense flare loops form in all the simulations including the $B_0=20$~G experiment that shows net downward mass flux through the upper layer of the chromosphere.  

There is some chromospheric evaporation in each simulation, despite the net downwards mass flux through the 5~Mm height plane in the simulation with $B_0=20$~G. The maximum velocity of these evaporations at heights of 5~Mm are shown in Fig.~\ref{fig:downshocks}f. These velocities scale from around 200~km s$^{-1}$ to around 500~km s$^{-1}$ over the simulations in reasonable agreement with typical values derived from ultra-violet line observations of chromospheric evaporations and 1D flare models \citep{2015Kennedy, 2016Polito, 2017Druett}. There are small peaks of high velocity that exceed the typically observed evaporation speeds in ultra-violet spectral lines, which reach up to nearly 800~km s$^{-1}$ in the experiment with $B_0=65$~G.

\subsection{Flare loop-tops} \label{sec:impactloops}

\begin{figure*}[!ht]
    \centering
    \includegraphics[width=0.8\textwidth]{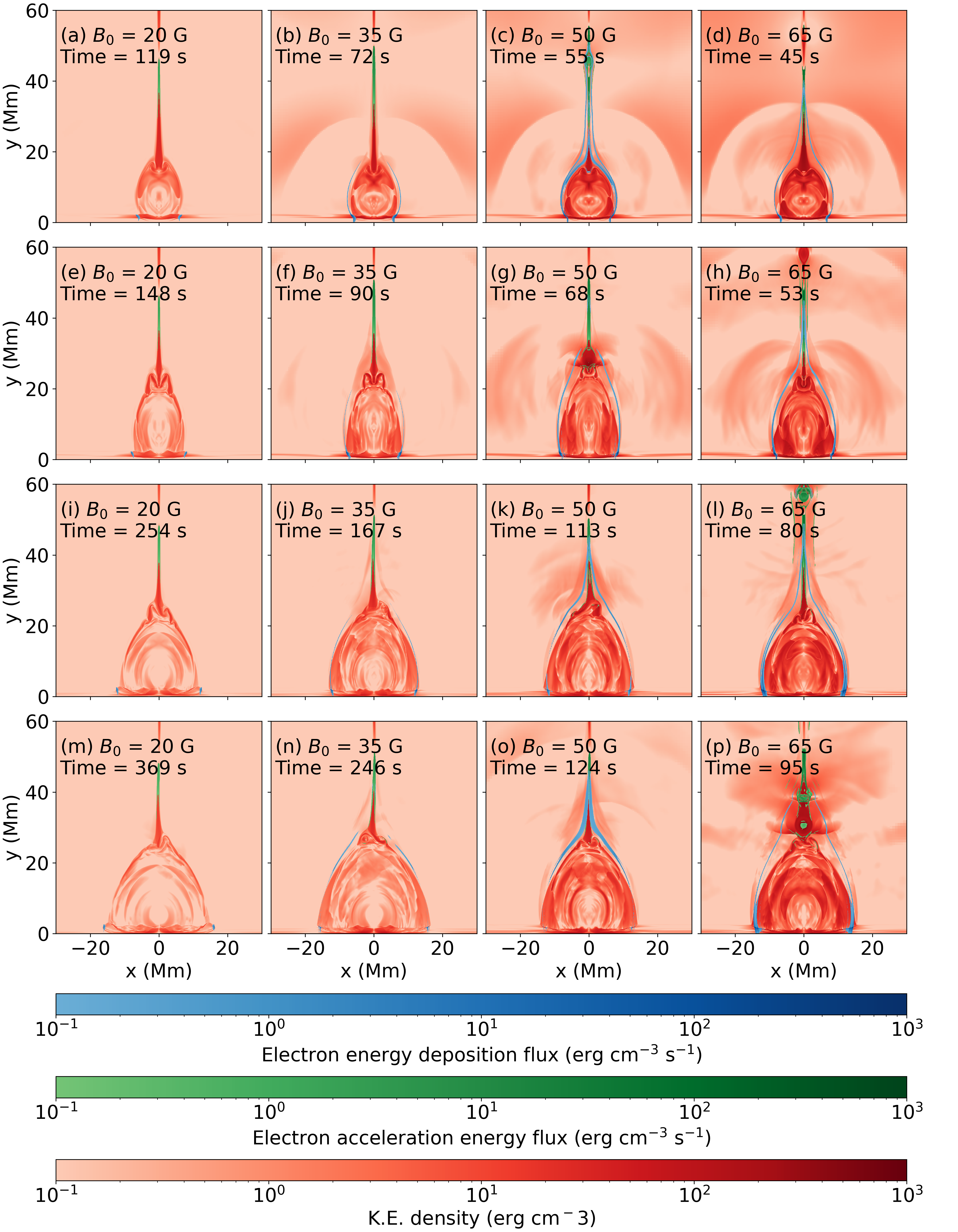}
    \caption{Kinetic energy density color maps. Kinetic energy density is shown in red and highlights the flare loops, the current sheets, and shocks travelling through the surrounding atmosphere. Overlays show the electron acceleration energy densities in green and the energy deposition density in blue.
    Energy densities for all panels of this figure have a lower saturation limit of $10^{-1}$~erg~cm$^{-2}$~s$^{-1}$, and an upper limit of $10^{3}$~erg~cm$^{-2}$~s$^{-1}$. The columns from left to right show results for the experiments with background magnetic field strengths of $B_0= 20$~G, $35$~G, $50$~G, and $65$~G respectively. The top row shows the simulations at the time after the impact of the leading edge of the reconnection out-flow jet. This impact and its reflection causes hot up-flows from the chromosphere. The second row shows the time at which the flare loops have rebounded after their compression during the impact. The third and fourth rows show later times when the loop-tops settle, and exhibit turbulent eddies on alternating sides of the central line, i.e. a magnetic tuning fork process. An animated version of the figure is provided in the online materials.}
    \label{fig:looptop}
\end{figure*}

Figure \ref{fig:looptop} presents kinetic energy density maps which highlight the current sheets and flare loops. We use them to illustrate the evolution of the flare loop-tops in the simulations with different magnetic field values. The top row shows the impact of the reconnection out-flow jets onto the chromosphere. Strong lateral flows away from the center of the impact move along the top of the chromosphere, and a shock-wave expands as a dome over the whole coronal domain outside of the flare loops, centered on those loops. This phenomenon is also present in the experiment with $B_0=20$~G but with kinetic energy densities values close to the lower saturation limit that make it hard to see in the figure. The impact and rebound of the reconnection out-flow on the lower atmosphere, as well as heat conduction, create strong evaporation flows up the flare loops from their footpoints. The densities and velocities of these evaporation flows scale with the density and velocity of the impacting reconnection out-flow jet, and thus with the background field strength as described in section \ref{sec:impactevap}. We note the absence of strong rebound up-flows in the $B_0=20$~G model, which is consistent with the lack of variation in chromospheric mass fraction and negative chromospheric mass flux for this model seen in Fig.~\ref{fig:downshocks}(d-e). Evaporation up-flows do also begin for the experiment with $B_0=20$~G at some time after the time shown in Fig.~\ref{fig:looptop}a, as can be seen in panel Fig.~\ref{fig:looptop}e. 

The down-flows from the loop-tops that are concurrent with the onset of chromospheric evaporation are slow shocks caused by the compression of the loop-top region during the impact, and the resulting negative pressure gradient outwards from the central position of the loop-top in the directions along the field-lines. 

The second row of diagrams shows the flare loop arcade after the rebound of the reconnection out-flow on the lower atmosphere. In each experiment the loop-tops display oblique and horizontal fast shocks, and potentially multiple termination shocks, as described in \citet{2015Takasao, 2016Takasao}. This pattern of shocks is a consistent feature across all the experiments. However, these typical behaviours as described in \citet{2015Takasao, 2016Takasao} can be disrupted by plasmoids which are ejected downwards from the current sheet in the experiments with stronger background field strength (see e.g. Fig.~\ref{fig:looptop}g, with $B_0=50$~G).

The magnetic tuning fork process \citep{2016Takasao} is a flare loop-top oscillation that is controlled by the back-flow of the reconnection out-flow. It is evident across all the experiments. This is shown in the lower two rows of panels, in which turbulent eddies form on each side of the termination shock, with a dominant extent that alternates between the left and right sides. These eddies are also associated with pulses of shocks that propagate out into the surrounding plasma and can be identified as sets of fringes in the kinetic energy density that move away from the loop-top locations in the lowest two rows of panels in Fig.\ref{fig:looptop}.

The magnetic tuning fork process and the plasmoids are both capable of sending high density flows outward around the turbulent loop-tops. Sometimes these flows are dense enough to intercept a significant fraction of the energy in the energetic beams of electrons before they can reach the chromosphere. This phenomenon can be seen in the video version of Fig.~\ref{fig:looptop} in which the beam energy deposition (blue colour) increases in the coronal region of the experiment (see $B_0=65$~G before and after $t=100$~s, $B_0=50$~G at around $t=123$ to $t=129$~s, and $B_0=35$~G after $t=200$~s). This interception of beam electrons is also visible in the chromospheric beam heating rates at the footpoints of the models in Fig.~\ref{fig:beamheat}. It cna be seen through the decreases in the electron beam energy reaching the chromosphere, due to the deposition of a significant portion of the beam fluxes in the corona.

In the $B_0=35$~G experiment we see a periodic brightening of the beam energy deposition rate in the coronal loops after $t=220$~s. This appears to be associated with the magnetic tuning fork process which periodically (around every 16~s) emits shocks into the surrounding loops. These shocks periodically increase the densities in field-lines that have recently reconnected, reaching maximum around the time that the tuning fork pulse passes from one side of the loop-top to the other, and occurring on both sides of the experiment. This increased density, in turn, removes energy from the beams of electrons before they reach the footpoints, with a periodicity of around 16~s. Faint traces of this process are visible in Fig.\ref{fig:looptop}n and the video versions of this figure. Thus the magnetic tuning fork can directly affect the loop-top X-ray emission as described in other papers \citep{2016Takasao, 2018McLaughlinQPP, 2021ZimovetsQPP}. For example, the waves emitted from the loop-top align with the concept of a periodic fast-mode magnetoacoustic wave as per the analysis of \citet{2016Takasao}, which used similar models to those presented here. It has been argued that such waves propagating towards reconnection x-points may also generate quasi-periodic-pulsations \citep[QPPs, ][]{2009McLaughlin}. Here we describe a multidimensional secondary process which can also explain simultaneous QPPs in footpoint and coronal loop HXR sources. Pure MHD models without beam electrons cannot self-consistently quantify these HXR or beam-related effects and their related to QPP variations. However, the lack of particle acceleration modelling for non-thermal electrons in the termination shock and turbulent reconnection loop-top regions means that the origins of QPPs in HXR sources cannot be definitively discerned from our models. In follow-up work we will produce a more rigorous analysis with direct synthetic images in wavebands relevant to these processes.

The experiment with $B_0=50$~G also exhibits periodic pulsing of beam energy deposition in the coronal loops, but in this and the $B_0=65$~G case the main factor in the disruption of the flow of electron beams from the X-point to the chromosphere is plasmoids. The reduction in beam electron footpoint heating in the case of $B_0=50$~G is evident at 120~s, just after a plasmoid strikes the loop-tops (Fig.~\ref{fig:beamheat}(c)), with the energy deposition around this time shown in Fig.~\ref{fig:looptop}(o). For the simulations with $B_0=65$~G plasmoids can be seen approaching the loop-tops at around 95-100~s in Fig.~\ref{fig:looptop}(p), and the subsequent reduction in beam heating of the chromospheric footpoints is evident for the rest of the experiment in Fig.~\ref{fig:beamheat}(d). 

Within a post-processing full kinetic model that was evolved over the backdrop of a resistive MHD-simulated atmosphere, \citet{2020KongAcceleration} found that the interaction of plasmoids with the termination shock in the loop-tops significantly modulated and softened (increased the negative exponent of the power law energy distribution) the electron beams accelerated. This attenuation and shift towards lower energy electrons was related to the collisions of the plasmoids with the looptops, because these collisions increased the number of grid cells showing compressions. Such modelling is highly valuable and may contribute to future works including recipes for the accelerated electron spectrum in longer, larger scale simulations at affordable computing costs. Our work does not include a detailed kinetic model for the accelerated spectrum of electrons, but the effect of the plasmoids is, rather coincidentally, similar for the thresholding of electrons reaching the chromosphere. In our experiment the collision of the plasmoid with the looptops causes density and magnetic field variations in the corona that result in both increased beam energy deposition near the termination shock (with reductions at the footpoints) and of greater particle trapping.

We generated SXR curves for each flare model (Not shown in figures) in Watts per metre squared, in order to produce GOES classifications. For this we use the method outlined in \citet{2018RuanKHI} based on the work of \citet{2015DelZanna, 2015Pinto, 2016Fang}. Their formula expresses fluxes in photon~cm$^{-2}$~s$^{-1}$. We adapt this by multiplying the integrand by the photon energy to produce a result in erg~cm$^{-2}$~s$^{-1}$. The integral is taken between limits with energies corresponding to those of the GOES 1-8~\AA\: band and then converted to W~m$^{-2}$. It is the peak of the flux in this GOES channel that defines the standard X-ray classification of a solar flare \citep{1970BakerGOESClassification}. These values need to be multiplied by a representative depth in 2.5D models. For consistency with the work of \citet{2023Ruan3D} we choose this depth to be 100~Mm. Assuming an arcade of this depth we obtain the data seen in table \ref{tab:flareclass}. The flare classifications are spread across a reasonable span of the observed range on the Sun, but do not reach the X-class flare classification.

\begin{table*}[!ht]
\begin{centering}
\begin{tabular}{ | c | c c c c | c c c | }
\hline
$B_0$ (G) &  $F_0$ class & max $F_0$ (erg cm$^{-2}$ s$^{-1}$) & time (s) & x (Mm) & GOES Class & max GOES 1-8\AA\: (W~m$^{-2}$) & time (s) \\ 
\hline
\textbf{20} & \textbf{1F9} & $1.40\times 10^{9}$ & 411 & 17.2 & \textbf{B1.5} & $1.53\times10^{-7}$ & 193 \\
\textbf{35} & \textbf{1F10} & $1.39\times 10^{10}$ & 237 & 15.4 & \textbf{C1.3} & $1.32\times10^{-6}$ & 123 \\
\textbf{50} & \textbf{4F10} & $3.57\times 10^{10}$ & 116 & 12.4 & \textbf{C5.5} & $5.53\times10^{-6}$ & 93 \\
\textbf{65} & \textbf{5F10} & $4.72\times 10^{10}$ & 79 & 11.6 & \textbf{M2.3} & $2.34\times10^{-5}$ & 113 \\
\hline
\end{tabular}
\caption{$F_0$ and GOES classifications of the simulated flares. The left column shows the background field strength of each simulation. The second gives the $F_0$ classification, and the third the flux of the electron energy deposition at its maximum value. The time and location of this maximum is shown in the 4th and 5th columns. The sixth column gives the GOES SXR classification as per the scheme of \citet{1970BakerGOESClassification}, as described in \citet{2022Pietrow}, and the seventh gives the peak of the flux in the GOES 1-8\AA\: channel, assuming a third dimension depth of 100~Mm, which scales the flux linearly. The final column shows the time at which this maximum GOES flux occurs.}
\label{tab:flareclass}
\end{centering}
\end{table*}

\subsection{Flows along a field-line} \label{sec:1D}

In section \ref{sec:beam}, statistics for the electron beam deposition were presented at each base point of the models. Now that we have presented the multi-dimensional evolution of the lower and upper atmosphere we inspect the variations in 1D of parameters along individual field-lines. Many of the physical processes in flare loop systems are field-aligned, and so there is significant value to inspecting the dynamics along such cuts. Moreover, this analysis provides a much greater basis than currently exists in the literature to enable the comparison of results of flare simulations in multiple dimensions with decades of research results derived from detailed 1D radiation hydrodynamic modelling of flare loops.

For this investigation, we inspect the strongest (M2 class) flare with $B_0 = 65$ G and a maximum beam electron energy flux over all space and time of $4.7\times 10^{10}$ erg cm$^{-2}$ s$^{-1}$. Field-lines with footpoints at $x=-10, -12.5$ and $-15$ Mm are selected as representatives of the variety of locations available within this multi-dimensional morphology. 

Plasma number densities, vertical velocities, temperatures, and kinetic energy densities are shown as functions of distance from the loop apex (y-axis) and time (x-axis) in Figs.~\ref{fig:field-line10}, \ref{fig:field-line12.5}, and \ref{fig:field-line15}. The electron fluxes along these field-lines are shown as functions of time, over-plotted as red lines on these images. These views form the direct counterpart of restricted 1D hydrodynamic models \citep[e.g. Fig.6 of][]{2023UnverferthPREFT}, and can be compared readily. 

Before magnetic reconnection the tracked field-lines exit the experiment at the top boundary, and after reconnection they reach to a footpoint on the opposite side of the flare loop system. Thus, there is a sharp disconnect between the top halves of these panels at times before and after the reconnection, as this portion of the field-line tracks completely different plasma. After the field-lines reconnect, they rapidly retract and shorten. This can be seen by the rapidly decreasing total length of the field-lines between the photospheric footpoints of the field-lines. The footpoints are plotted as green lines at the tops and bottoms of the panels.

\subsubsection{$B_0=65$ G, $x=-10$ Mm, $F0=$1.0F10} \label{sec:x10}

\begin{figure*}
    \centering
    \includegraphics[width=0.9\textwidth]{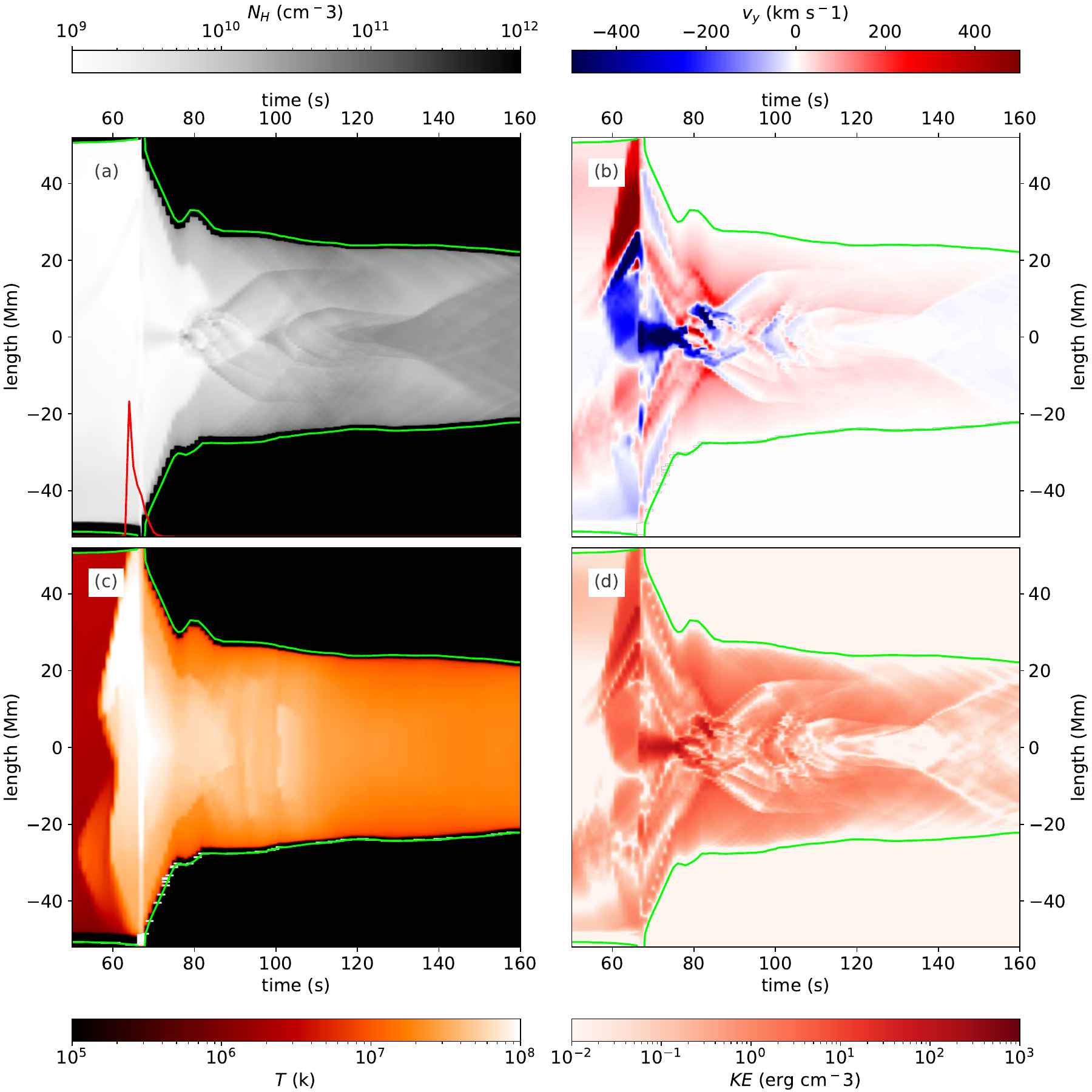}
    \caption{Evolution of the atmosphere along a single field-line with one footpoint at $x=-10$ Mm, discussed in section \ref{sec:x10} ($B_0=65$ G, $x=-10$ Mm, $F0=$1.0F10). These are shown in plots of time on the horizontal axis, and length $s$ along the field-line vertically, with $s=0$ at $x=-10$ Mm, $y=0$ Mm. The parameters shown are (a) plasma number density, (b) the vertical velocity, (c) the plasma temperature (d) the kinetic energy density. The plasma number density panel show the beam electron energy flux deposited in the chromosphere above the left footpoint, over-plotted in red. This over-plot is scaled such that the maximum electron energy flux deposited throughout the entire simulation ($4.7\times 10^{10}$ erg cm$^{-2}$ s$^{-1}$) corresponds to the peak reaching top of the panel, with zero at the bottom. The beam in this field line reaches a peak flux of $1.0\times 10^{10}$ erg cm$^{-2}$ s$^{-1}$. Note that the field line changes in overall length as function of time. The extent of the experimental domain is highlighted with green lines in each panel. Before reconnection this highlights the bottom and top of the experiment, after reconnection these highlight the locations of the photospheric footpoints in each time-step. Values outside of this are saturated to their photospheric values for continuity, but do not represent simulated values.}
    \label{fig:field-line10}
\end{figure*}

From figure~\ref{fig:field-line10} one can see that this field-line reconnects at $t=67$~s. This is $4$~s after $t=63$~s, when the beam electrons switch on for this field line, accelerated in regions away from the x-point. The onset of the evaporation can be seen from the lower footpoint at $t=65$ s. This requires some explanation with respect to the standard flare model. Before reconnection there is significant heating (to around $T=10$ MK) on this field-line from energy supplied by neighbouring plasma via methods including heat diffusion and the expansion of the neighbouring flux tubes which generates compressive heating of the flux tube we are inspecting. The heating intrudes at lengths/heights around $s=-25$ Mm at $t=53$ s. This causes upward and downward flows to expand outward from this point. More dramatic heating occurs at $t=60$~s just before the reconnection time, both at the similar heights the previous heating and near the reconnection region higher up the open field-line. This heating results in heat conduction and velocity flows. It is this conduction front approaching the chromosphere which drives evaporation up from the footpoints. The evaporated material continues to be heated and expand, driving further acceleration up to around 300 km s$^{-1}$ by the time the evaporation reaches $y=20$ or $s=-10$ Mm, at temperatures of around 2 MK. 

Meanwhile, the top of the reconnected loop collapses downwards at high velocity, and shortens in total length. This process is visible as a dark blue horizontal stripe in vertical velocity (down-flow) immediately after the reconnection event, seen in panel (b). The apex of the line shortens until it is around $s=20$ to $s=30$ Mm away from the footpoints. The compression of the loop-tops drives a series of hot (50-100 MK) outward (downward) flows from the apex, starting at around $t=80$ s. The velocity plot, kinetic energy plot, and temperature structure of the loop-tops show that the region is undergoing turbulence as well as heating events. 

The downward flows driven from the loop-tops meet the rising chromospheric evaporation. The evaporation and downward travelling loop-top flows shock when they meet, reducing the velocities of both streams and heating the upward moving material significantly. This can be seen in the changes of gradients of the rising and falling density fronts in figure~\ref{fig:field-line10}a at around $s=\pm 8$ Mm, at $t=80$ s. In the beam driven evaporation model of \citet{2023DruettEvap} there was no strong reconnection out-flow to compress the loop tops and drive downward flow that meet the rising evaporation, in that study a significant portion of the evaporated plasma passed over the loop-tops and down to the other side of the arcade. In contrast, in this study the upward evaporations do not directly pass over the loop-top region, but get caught in the turbulence until gentle down-flows form at around $t=140$ s. The direct passing of up-flows over the loop apex and down towards the opposite footpoint can also be seen in 1D loop models such as in the central and right panels of Fig 6 in \citet{2023UnverferthPREFT} which also shows no loop-top turbulence, due to the 1D nature of the model. 

Along this field-line, pulses of extra density and kinetic energy rise upwards from both footpoints after around $t=85$ s which become broader, slower and weaker over time. This occurs after the beam electron processes have ceased along the field-line, possibly indicating some wave-like behaviour.

\subsubsection{$B_0=65$ G, $x=-12.5$ Mm, $F0=$2.5F10} \label{sec:x12.5}

\begin{figure*}[!ht]
    \centering
    \includegraphics[width=0.9\textwidth]{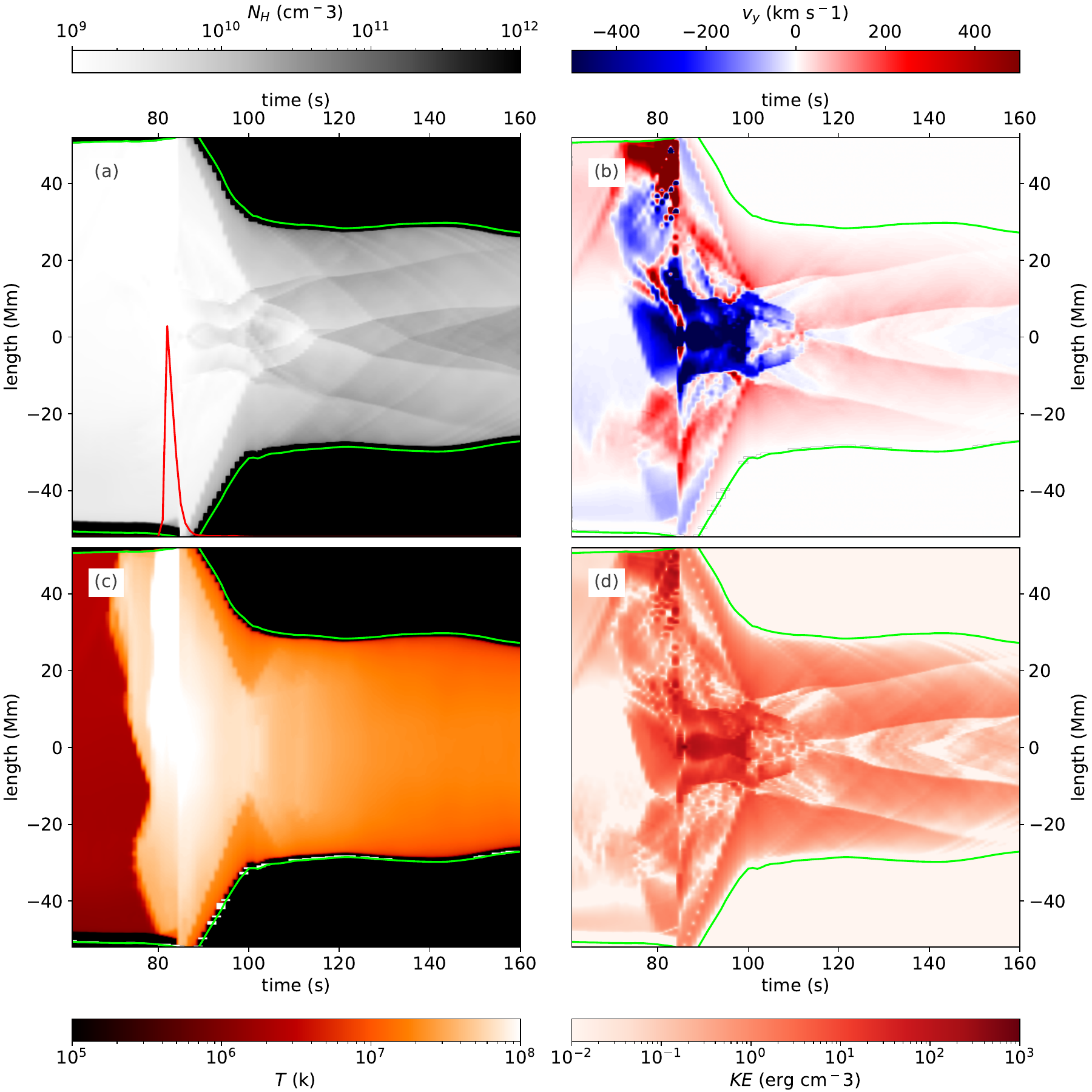}
    \caption{Evolution of the atmosphere along a single field-line with one footpoint at $x=-12.5$ Mm, discussed in section \ref{sec:x12.5} ($B_0=65$ G, $x=-12.5$ Mm, $F0=$2.5F10). The formatting is the same as that used in Fig. \ref{fig:field-line10}. The beam in this field line reaches a peak flux of $2.5\times 10^{10}$ erg cm$^{-2}$ s$^{-1}$.}
    \label{fig:field-line12.5}
\end{figure*}

Figure~\ref{fig:field-line12.5} shows that, predictably, the field line further out reconnects later ($t=85$ s). Both the beam acceleration and the evaporation processes from the footpoints start at 80~s. This is also co-temporal with the arrival of the fast downward propagating hot jet due to heating and expansion of neighbouring material that heats this flux tube around $s=-25$ Mm beginning at around $t=75$ s. The driving of the evaporation is broadly similar to that described for the footpoint at $x=-10$ Mm. After reconnection the loop-top collapses downwards at greater velocities, and halts with a wider span of $s$ values. The simulation ends before the gentle down-flows from the loop-tops can reach the footpoints.

\subsubsection{$B_0=65$ G, $x=-15$ Mm, $F0=$5.7F9} \label{sec:x15}

\begin{figure*}[!ht]
    \centering
    \includegraphics[width=0.9\textwidth]{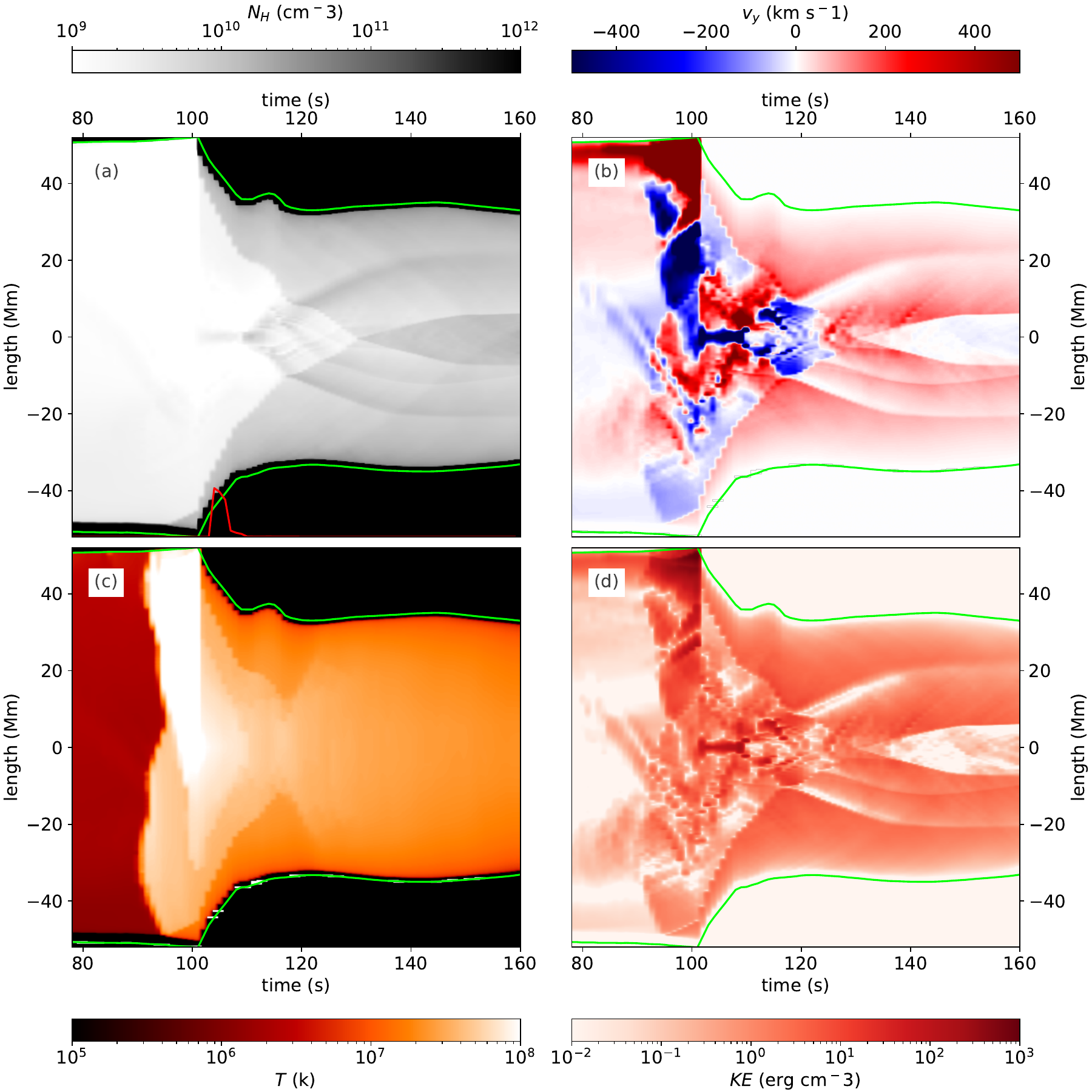}
    \caption{Evolution of the atmosphere along a single field-line with one footpoint at $x=-15$ Mm, discussed in section \ref{sec:x15} ($B_0=65$ G, $x=-15$ Mm, $F0=$5.7F9). The formatting is the same as that used in Fig. \ref{fig:field-line10}. The beam in this field line reaches a peak flux of $5.7\times 10^{9}$ erg cm$^{-2}$ s$^{-1}$.}
    \label{fig:field-line15}
\end{figure*}

Figure~\ref{fig:field-line15} shows that this field-line experiences chromospheric evaporation (at $t=94$~s) well before reconnecting (at $t=102$~s), and before the beam electrons reach the chromosphere (at $t=104$~s). The heat driven expansion from the nearby loops begins at positions near $s=-25$ Mm. Again, the hot plasma expands in upwards and downwards directions from $s=-25$ Mm. This flow reaches the chromosphere at $t=93$~s, and immediately drives chromospheric evaporation which achieves similar speeds of around 300 km s$^{-1}$. The evaporation collides with down-flows at positions of around $s=20$~Mm. 

For this field-line which is situated in the outer region of the flare, by the time it reconnects, significant loop-top turbulence is already present. This can be seen through the high-speed, direction-varying velocities in the loop-tops (panel b). Also, before the time of the reconnection of the field-line there are already some slightly higher-density features in the loop tops, as well as the rising chromospheric evaporation fronts. The beam electrons deposit a significant fraction of their energy in these evaporation and loop-top features. Thus the beam flux reaching the chromosphere is significantly reduced. A dedicated future investigation will be necessary to understand these effect in detail, including in experiments with evaporation driven by beam electrons \citep{2023DruettEvap}.

\section{Summary and discussion} \label{sec:summary}

In this paper we presented a study of the 2.5~D \texttt{MPI-AMRVAC} flares including beam electrons, first reported in \citet{2020RuanFlare}. 
By varying the background magnetic field strength by a factor of 3.25 in these simulations between $B_0 = 20$~G and $65$~G, the GOES classification of the simulation changes by over 2 orders of magnitude, between B1.5 and M2.3 (assuming a 100~Mm arcade length in the third dimension, see table \ref{tab:flareclass}). 

The flux of energy supplied by energetic electrons at any given chomospheric footpoint has a characteristic duration of between 5 and 20 seconds, usually with a relatively triangular profile of flux against time, peaking earlier in the profile (see Figs.~\ref{fig:beamheat}, \ref{fig:field-line10}, \ref{fig:field-line12.5}, and \ref{fig:field-line15}). The peak beam heating flux at a footpoint in each experiment varies between $1F9$ for the case $B_0 = 20$~G ($F_0=1\times 10^{9}$ erg cm$^{-2}$ s$^{-1}$) and $5F10$ for the case $B_0 = 65$~G ($F_0=5\times 10^{10}$ erg cm$^{-2}$ s$^{-1}$), over and order of magnitude difference. This is the first paper reporting the details of chromospheric beam fluxes and their evolution in multi-dimensional simulations.

In all simulations, bi-directional reconnection out-flow jets are formed in the corona at heights of 50~Mm where the initial reconnection x-point forms. The out-flows have a classic "lobster claw" shape \citep{2011_Zenitani_reconnection_outflows}. A fast shock exists in the tail of this feature and stabilises some time after the out-flow impacts the lower atmosphere. The maximum speed achieved in these flows scales by a similar amount to the background field, from around 1000 km s$^{-1}$ to 3200 km s$^{-1}$ across the experiments with $B_0 = 20$~G to $B_0 = 65$~G respectively. As a result, after the loop system settles, the maximum out-flow speed is approximately constant across the simulations when expressed as an Alfv\'en mach number (see Fig.~\ref{fig:outlfowsalfven}). It is possible that this ${\cal{O}}(10)$ maximum value would alter based on the variation of other simulation parameters, such as the vertical position of the resistivity patch, which would provide a longer or shorter reconnection out-flow jet if placed higher or lower in the atmosphere respectively. The maximum out-flow Alfv\'en mach number was insensitive to changes of the maximum anomalous resistivity.

We perform the first detailed investigation of chromospheric response to the impacts of reconnection out-flow jets in multi-dimensional models, including the changes in these responses across a variety of flare strengths. The impact of the reconnection out-flow jets and the heat conduction front on the lower atmosphere heats the chromospheric material. This generates hot up-flows ($T \sim 2$ MK, chromospheric evaporation, see Figs.\ref{fig:field-line10} to \ref{fig:field-line15}, panel c) and cooler down-flows ($T \sim 20,000$ K, "down-flowing chromospheric compressions", see Figs. \ref{fig:chromo80} and \ref{fig:chromo_late}). 
Chromospheric material is also heated from around 6000~K to temperatures around 20,000~K within a second of the beam electrons being switched on. Heating of the plasma to temperatures around 20,000~K extends downwards from the tops of the chromosphere to depths of 1.5~Mm for the $B_0 = 20$~G simulation and to 1.0~Mm for the $B_0 = 65$~G case. There are noticeable kinetic energy imprints of the beam electrons at the chromospheric footpoints of the flares after the acceleration is switched on, but these are swamped by the reconnection out-flow jet in the present simulation suite (see Fig.~\ref{fig:beamKEsignatures}). This is investigated in more detail in a companion paper \citep{2023DruettEvap}.

At heights of 300~km above the photosphere the downward kinetic energy flux densities reach $5\times 10^6$ erg cm$^{-2}$ s$^{-1}$ for the experiment with $B_0 = 20$~G, and up to $4\times 10^8$ erg cm$^{-2}$ s$^{-1}$ for $B_0 = 65$~G. This demonstrates a significant transfer of energy and momentum to the photosphere, however even the "down-flowing chromospheric compression" for the strongest flare presented drops below the local sound speed at heights between 200 and 300~km above the photosphere. This implies that we would not expect these simulated flares to produce sunquakes via the "down-flowing chromospheric compressions". This is the first set of multi-dimensional flare simulations to test down-flowing chromospheric compressions as potential drivers of sunquakes (see \citet{2016Russell} for an investigation in multiple dimensions that was restricted to the lower atmosphere). However, the lower atmospheres of our simulations are simplified, with field strengths and densities that are significantly lower at the base of the model than those considered to be typically "photospheric", and this will be addressed in a future study.

Regarding the excavation of the chromosphere by evaporation processes in the flares, our 2.5D simulations (Figs.~\ref{fig:chromo80}, \ref{fig:chromo_late}) bear a visual resemblance to the ribbon height substructure that can be seen in observations of the chromosphere using COCOPLOTs \citep{druett_cocoplot_2022, 2024PietrowRibbon}, with flare ribbon evolution described in \citet{2023PietrowHARPS}. In these observations the flare ribbon emission appears to be coming from lower heights than the chromospheric emission from just outside the boundaries of the flare ribbons. This can be inferred from the projection effect of the cooler chromospheric material outside the flare ribbon, which is overlapping the adjacent bright flare ribbon emission in the line of sight, leading to strong absorption of the flare emission. Moreover, this effect is clearly present on the leading edge of the eastern ribbon in that paper which, due to the viewing angle, is oriented such that if the ribbon formation was lower than that of the surrounding chromosphere, should be overlapping. Structures much more similar to the ribbon substructures reported by Singh et al. (private communication) may relate to the periodic evaporation pulsations noted in our simulations (Figs.~\ref{fig:field-line10} to \ref{fig:field-line15}). Between the ribbons there appears to be a higher plateau near the polarity inversion line that is also reproduced by the combination of the out-flow impacts and the magnetic topology of the reconnected field lines in our experiments. In a future work we will follow-up these MHD plus beam driven runs with non-LTE spectroscopic analysis, now possible for multi-dimensional setups.

The heat conduction, impact of the reconnection jets, and beam heating of the lower atmosphere (principally the impact and heat conduction) drive chromospheric evaporation with characteristic speeds at a height of 5 Mm ranging from $200$ km s$^{-1}$ to $600$ km s$^{-1}$ across the range of the background field strengths studies, again scaling relatively linearly with this parameter. The maximum at any time in the $B_0=65$~G experiment is $\sim800$~km~s$^{-1}$, which is higher than typically observed speeds. 

Notably, there is a downward net mass flux through a height of 5~Mm in the $B_0=20$~G experiment, indicating that more mass is ejected downward along the coronal loops due to the reconnection than is evaporated upward from the surface due to the energy transport. For all the other (stronger) flares there were positive mass fluxes at this height, and a clear reduction in the proportion of cool chromospheric material at low atmospheric heights due to the plasma heating and evaporation (Fig.~\ref{fig:downshocks}).

The upward evaporation from the chromosphere meets fronts travelling in the opposite direction, down from the loop-tops. These downward fronts are squeezed by the pressure gradient force along the field-lines due to the impact of the reconnection out-flow jet on the loop-tops. These two fronts shock and reduce in speed when they meet (Figs.~\ref{fig:field-line10}, \ref{fig:field-line12.5}, and \ref{fig:field-line15}). The evaporation does not directly travel from footpoint to footpoint in contrast to a similar experiment with chromospheric evaporation driven by beam electrons \citep{2023DruettEvap}. This difference could prove highly instructive in discerning flare evaporation mechanisms if it persists across robust variations of the simulation parameters. 

The evolution of the horizontal and oblique shocks in each experiment is similar to the descriptions of \citet{2015Takasao, 2016Takasao} across all experiments. Also turbulent vortices form on alternating sides of the loop-tops with time. This creates the magnetic tuning fork phenomenon \citep{2016Takasao} which has been suggested as a candidate process for producing flare quasi-periodic pulsations in loop-top emissions \citep{2018McLaughlinQPP, 2021ZimovetsQPP}. 

We propose a new mechanism for generating simultaneous QPPs in the footpoint and loop top HXR sources. The magnetic tuning fork process produces pulsations in the SXR loop-top sources \citep{2016Takasao}. In our simulations these will have similar periods to pulsations in the loop top and footpoint HXR bremsstrahlung sources. This is because the magnetic tuning fork process contributes to the creation of periodic variations in the densities of material along the recently reconnected loops. This, in turn, attenuates the fluxes of electrons that reach the chromosphere from acceleration locations above the loop-tops. Imaging X-ray spectra would be necessary to observe this effect in the HXR foot-point sources, which could be provided by the proposed FOXSI instrument \citep{2023FOXSI}. If in a future study, line synthesis of chromospheric spectral lines in our simulations can be achieved, then we would be able to infer whether this process would also produce pulsations in visible and near-UV sources.

Our investigation of the flows along 1D field-lines reveal several multi-dimensional effects that are not accounted for in 1D studies of solar flare loops, even those attempting to recreate multi-dimensional effects such as \cite{2020Kerr1D3D, 2023UnverferthPREFT}. Firstly, the reconnection and loop top turbulence sources are intimately linked to the multi-dimensional nature of the simulation.
Secondly, a shock occurs when chromospheric evaporation meets downward loop-top sources that have been forced along the field-lines due to the high pressure in the loop tops. Thirdly, there are sources of chromospheric evaporation caused by heat conduction via field-aligned transport from the loop-top sources, but also from neighbouring flux tubes via processes including heat diffusion and compression. In strong flares these processes cause the leading edges of the flare ribbons to begin evaporating material by thermal conduction before the associated field lines have reconnected. These regions completely lack beam electrons and HXR sources at these times (Fig.~\ref{fig:field-line15}). This matches the pattern of up-flows from ultraviolet satellite observations by the Interface Region Imaging Spectrometer \citep[IRIS, ][]{iris_2014}, as reported in \citet{polito_ribbons_2023}, who note that the leading edges of the flare ribbons show significantly lower evaporation of chromospheric material than the main bodies of the flare ribbons. \citet{polito_ribbons_2023} interpret that the beams of electrons in the leading edges of the flare ribbon are significantly weaker (presumably associated with acceleration processes near the X-point, in the current sheet above the flare loops) than those beams inside the body of the flare ribbon (which may be associated with acceleration sources nearer the termination shock, the loop-top turbulence, or the flare loop-tops themselves). Our models demonstrate that additional and alternative interpretations should be considered to those that can be provided by 1D analysis. However, modelling advances are required in our simulations to include particle acceleration in the termination shock and turbulent loop-tops before we can provide a comprehensive answer. 

Finally, we note a number of improvements that can be made to these models and the benefits these would bring. This lineage of simulations has focused on accurately reproducing coronal conditions, and future simulations should improve the lower atmosphere by increasing the magnetic field strength (eventually by several orders of magnitude at the base), and making the density profile accurately represent chromospheric and photospheric values. This would improve the accuracy and credibility of interpretations derived from spectral line synthesis regarding the motions of the "down-flowing chromospheric compressions", ribbon formation, and evaporation processes, thereby better constraining the energetics and fundamental flare processes responsible for visible, UV and SXR emissions. 

The beam model should be improved so that it can self-consistently be the principle agent driving evaporation. This has recently been achieved for the first time in multiple dimensions in a companion paper, \citet{2023DruettEvap}. The energy spectrum and mean pitch angles of these beams can be parameterised based on atmospheric quantities and acceleration statistics from detailed studies, including approaches by \citet{2018Bakke, 2020Frogner, 2023Frogner}. Effects such as self-induced electric field and return currents could also be included in the transport model \citep{1995Zharkova}. Energetic protons could also be considered in the particle transport modelling \citep{2015Zharkova}.

Detailed non-Local Thermodynamic Equilibrium radiative transfer \citep{2005Allred, 2012Carlsson, 2015Allred, 2022Hong} and non-equilibrium ionisation \citep{2007Leenaarts} can be included in the lower atmosphere. This could be combined with optically thick spectral line synthesis \citep{2021OsborneLIGHTWEAVER, 2023Jenkins}. The combination of these advances applied to our models would provide multi-dimensional context to our understanding of inhomogeneities, flows, energy delivery, and ribbon substructures observed in a large number of currently debated flare phenomena based on observations in the visible and near ultraviolet emissions of flares \citep{1984Ichimoto, 2022Osborne, 2022Pietrow, polito_ribbons_2023}, including investigating highly broadened flare ribbon spectral line profiles \citep{2020Zharkov, DruettPoster, 2022KowalskiBroad, 2024KerrBroad}, and the mechanisms responsible for sunquakes \citep{2001Kosovichev_sunquake, 2019Quinn_Sunquake}.

Further studies using this model would be instructive. 
An investigation of the effects of varying the height of the resistivity patch and introducing asymmetries into the simulation would examine the robustness of relationships derived using our standard set-up.
Data driven simulations and a systematic comparison of our multi-dimensional simulations with 1D simulations that include detailed physics would allow us to determine the most potent admixture of these approaches to use in future studies.
Furthermore, an equivalent study for simulations with evaporation driven by beam electrons, and a 3D version of the simulation would allow us to interpret observational signatures of the different evaporation mechanisms, thereby determining which processes principally drive evaporation and other fundamental flare phenomena on the Sun.

\appendix

\section{Beam heating and approximation of energy conservation} \label{sec:beam_heat_appendix}

\begin{figure}
    \centering
    \includegraphics[width=0.45\textwidth]{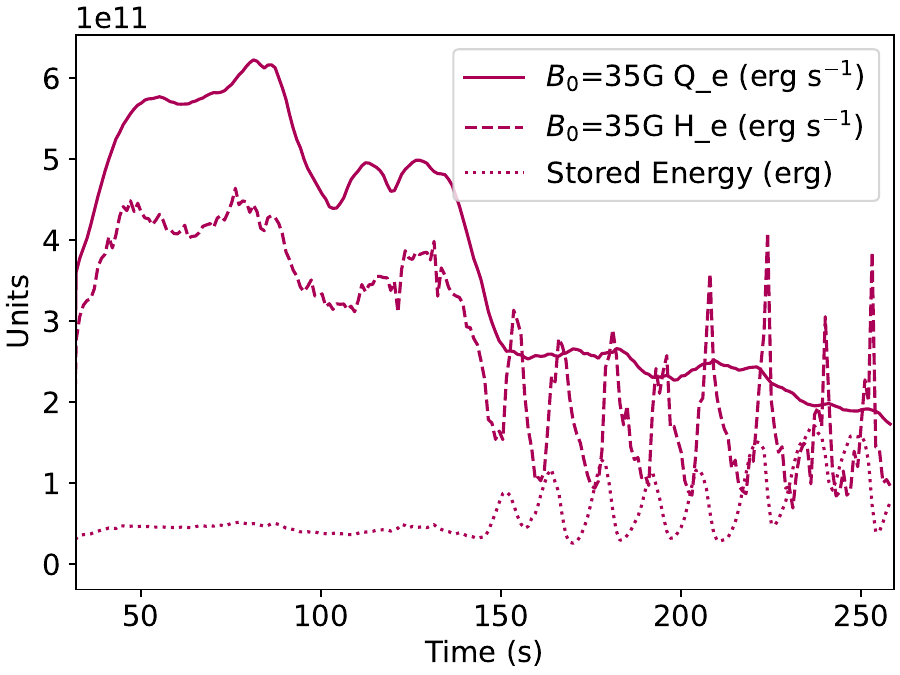}
    \\
    \includegraphics[width=0.45\textwidth]{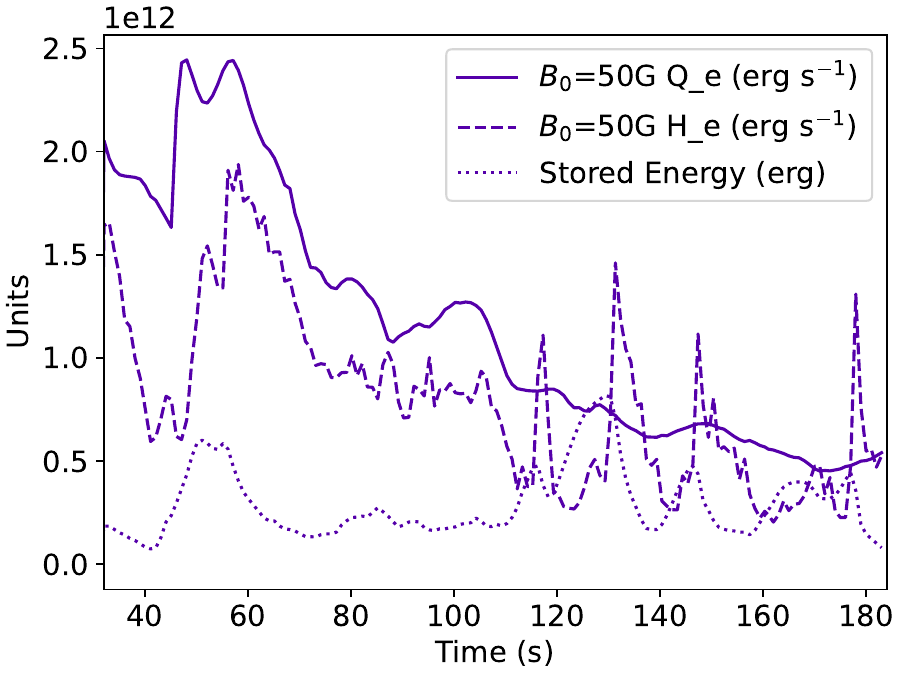}
    \caption{The approximate energy conservation of the beam processes. Results are shown for the $B_0=35$~G and $B_0=50$~G simulations assuming an arcade depth (3rd dimension depth) of 100~Mm. The solid line shows the Ohmic heating rate $Q_e$ (integrated over the whole domain) that is subtracted from the MHD model as per Equation \ref{eq:Q_e} in units of erg~s$^{-1}$. The dashed line shows the beam heating rate (re-interpolated onto the adaptive mesh grid and integrated over the whole domain) that is the source term $H_e$ in the MHD energy equation in units of erg~s$^{-1}$. The dotted line shows, at each time, the total energy stored in the field-lines that is carried over from the previous time-step, in units of erg.}
    \label{fig:beam_energy_cons}
\end{figure}

Once the field-lines that host energetic electron beams have been identified within the multi-dimensional simulation as described in section \ref{sec:beamdesc}, the 1D model of \citet{1978Emslie} is applied to the density profiles along their paths, creating a table of energy deposition rates along each section of each field-line. An interpolation routine is then used to redistribute the energy deposited from the beam back into the plasma \citep[see][Appendix D]{2020RuanFlare}. % update_heating_table, get_heating_rate, get_Qe

Electrons are considered trapped on the fieldlines when the atmosphere through which they are travelling is such that the beam electrons gain a perpendicular pitch angle to the field line. This is modelled by consideration of the first adiabatic invariant, which produces the following relationship \citep[][Appendix A]{2020RuanFlare} between the cosine of the electron beam's mean pitch angle, $\mu$, at position $s$ and its initial value at the top of the loop, $\mu_0$. This is expressed in terms of the initial field strength $B_0$ and the field strength at position $s$,
\begin{equation}
    \cos(\theta) = \mu(s,t) = \sqrt{1-\frac{B(s,t)}{B_0(t)}(1-\mu_0^2(t))} \le 0 .
        \label{eq:mu}
\end{equation}
Thus in Equation \ref{eq:mu}, if there exists any point $s$ along the field-line where the term that is inside the square root becomes negative, the heating rate is truncated at this position and the remaining energy stored in the electrons is considered to be trapped, and carried over to the next time-step. The energy and mean pitch angle of the electrons in the new time-step is found including considering contributions from both the newly generated and the carried-over electron distributions. Any electrons reaching the base of the model leave the system and take with them any remaining energy.

Absolute energy conservation is not guaranteed by the interpolation of energy deposition from the 1D models back into the multi-dimensional domain. This applies to the beam electron energy as it deposits energy back into the MHD plasma. Energetic electrons may also exit the domain and thus make the process an overall sink of energy. A comparison was made between the energy removed from the MHD equations at the acceleration sites, $Q_e$, and that returned to the plasma, $H_e$ (Fig.\ref{fig:beam_energy_cons}). The energy removed and returned track each other closely in form, and the total energy returned to the plasma never exceeds the total energy removed from the plasma. This holds true of the integrated totals at all times. Energy stored in field-lines by particle trapping in previous time-steps can be returned swiftly to the plasma and cause a transient excess in the heating rate, $H_e$, over $Q_e$. This excess energy is drained from the reservoir of the energy stored in the field-lines. The strong periodicity of this process shown at times after $t=150$~s in the top panel of Fig.\ref{fig:beam_energy_cons} is related to the processes described in section \ref{sec:impactloops}.

\section{Relative importance of the beam heating in the evaporation scheme}

\begin{figure}
    \centering
    \includegraphics[width=0.45\textwidth]{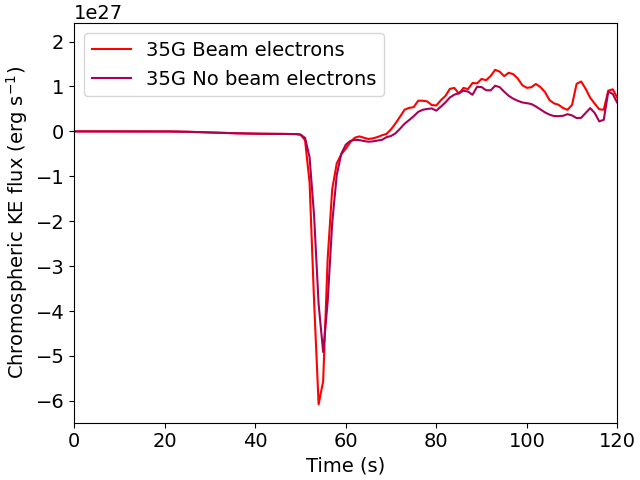}
    \\
    \includegraphics[width=0.45\textwidth]{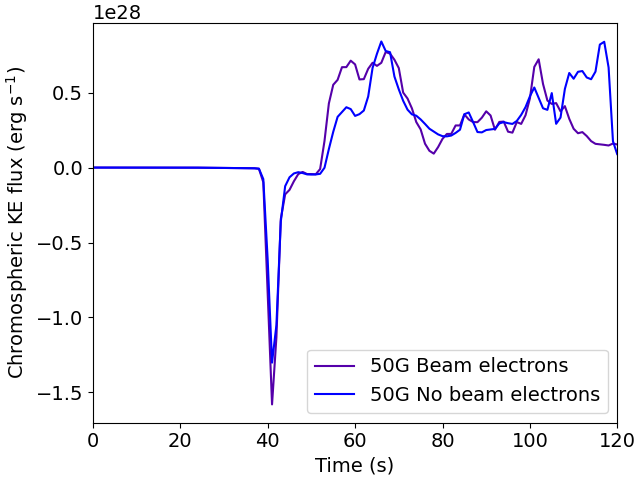}
    \caption{Kinetic energy fluxes through a height of 5~Mm in the simulations, as functions of time. These are shown for the $B_0=35$~G and $B_0=50$~G experiments in the top and bottom panels respectively assuming an arcade depth (3rd dimension depth) of 100~Mm. Results are shown for the main simulation presented in this paper alongside results for a simulation with identical parameters, except for setting the beam heating term, $H_e = 0$ at all times.}
    \label{fig:no_beam_heating}
\end{figure}

Figure \ref{fig:no_beam_heating} demonstrates the relatively minor contribution of the beam heating to the overall evaporation scheme by plotting the kinetic energy fluxes through a height of 5~Mm for experiments with and without the beam electron heating term, $H_e$, activated. The beam electron acceleration term, $Q_e$, was active in all cases so that the coronal x-point evolution was as near-identical as possible. These figures can be directly compared with the kinetic energy fluxes from the top panel of figure 5 in \citet{2023DruettEvap} where the evaporation was significantly driven by the beam electrons. In the simulations presented in this work heat conduction and the impact and rebound of the reconnection outflow can be seen to be by far the greatest influences on the evaporation and downwards travelling kinetic energy signatures.

\begin{acknowledgements}
%Personal
MD is supported by FWO project G0B4521N.
WR was supported by a postdoctoral mandate (PDMT1/21/027) by KU Leuven. 
RK is supported by Internal Funds KU Leuven through the project C14/19/089 TRACESpace and an FWO project G0B4521N. 
MD, WR, and RK acknowledge funding from the European Research Council (ERC) under the European Union Horizon 2020 research and innovation program (grant agreement No. 833251 PROMINENT ERC-ADG 2018). The computational resources and services used in this work were provided by the VSC (Flemish Supercomputer Center), funded by the Research Foundation Flanders (FWO) and the Flemish Government, department EWI.
MD acknowledges fruitful discussions with Alexander G.M. Pietrow regarding the contents of this manuscript.
\end{acknowledgements}

\bibliographystyle{aa}
\bibliography{refbib}

%\begin{appendix}
%\end{appendix}
\end{document}